\documentclass[fleqn,10pt]{wlscirep}
\usepackage[utf8]{inputenc}
\usepackage[T1]{fontenc}
\title{Coupling nanoscopic tomography and micromagnetic modelling to assess the stability of geomagnetic recorders}

\author[1,2,*]{Ualisson Donardelli Bellon}
\author[2,*]{Wyn Williams}
\author[1]{Ricardo Ivan Ferreira Trindade}
\author[3]{Ana Diaz}
\author[4]{Douglas Galante}
\affil[1]{University of São Paulo, Institute of Astronomy, Geophysics and Atmospheric Sciences (IAG), Department of Geophysics, São Paulo, 05360020, Brazil}
\affil[2]{University of Edinburgh, School of Geosciences, Edinburgh, EH9 3FE, Scotland}
\affil[3]{Paul Scherrer Institute, 5232 Villigen PSI, Switzerland}
\affil[4]{University of São Paulo, Institute of Geosciences, Department of Sedimentary and Environmental Geology, São Paulo, 05508080, Brazil}
\affil[*]{Correspondent authors: ualisson.bellon@usp.br, wyn.williams@ed.ac.uk }

\begin{abstract}

The recording of planetary magnetic fields is often attributed to uniformly-magnetised nanoscopic iron oxides, called single-domain (SD). Yet, the main magnetic constituents of rocks are more complex, non-uniformly magnetised grains in single or multi-vortex states. We know little about their behaviour due to limitations in defining their precise shape and internal magnetic structure. We propose a novel approach combining non-destructive synchrotron-based ptychographic nanotomography with micromagnetic modelling to explore the magnetic stability of remanence-bearing minerals. Applied to a microscopic rock sample, we identified hundreds of nanoscopic grains of magnetite/maghemite with diverse morphologies. For some grains, shape irregularities near the transition from SD to the single-vortex state allow for multiple domain states, some unstable and unable to record the field for significant periods. Additionally, some other grains exhibit temperature-dependent occupancy probabilities, potentially hampering experiments to recover the intensity of past magnetic fields. 

\


\

\end{abstract}
\begin{document}

\flushbottom
\maketitle

\section*{Introduction}

Measuring the magnetic properties and identifying the magnetic minerals in rocks is essential to palaeomagnetism, which studies Earth's magnetic field strength and direction throughout the planet's geological history \cite{Tauxe-2010}. While directional palaeomagnetic data is the only quantitative measurement that allows geoscientists to study palaeoconfigurations of ancient continental landmasses \cite{Merdith-2021}, databases of the past magnetic field strength (palaeointensity)  \cite{Bono-2022} provide links to the complex magnetohydrodynamic processes occurring deep in Earth's core \cite{Labrosse-2003, Driscoll-2016, Zhou-2022}. Quantifying the magnetic phases in rocks also provides essential information on globally scaled geological/climate events \cite{Liu-2012}. Rocks usually contain a mixture of magnetic particles in various sizes, including very stable nanoscopic uniformly magnetised grains (said to be in a single-domain (SD) state) and larger non-uniformly magnetised grains, either in pseudo-single-domain (PSD) or in multi-domain (MD) states. The palaeomagnetic theory is mainly based on Néel\cite{Neel-1955} ferromagnetic theory for SD grains, which relies on the capacity of these particles to retain their original remanence for billions of years but accounts only for a small proportion of magnetic particles contained within a typical palaeomagnetic sample.

Palaeomagnetic analysis of a samples' magnetic mineralogy and stability is usually based on bulk magnetic measurements, often capable of detecting trace amounts of magnetic minerals. Although the data obtained through these analyses use statistical procedures to ensure their reliability, critical magnetic phenomena with geological/geophysical implications can sometimes be masked in bulk magnetic measurements. The complexity of natural geological materials and the diversity of composition/geometry of natural remanence carriers in rocks indicate that the conditions assumed by classical palaeomagnetic experiments can be far from ideal. A detailed understanding of the magnetic recording reliability requires a nanoscale interpretation of the magnetic fidelity of magnetic mineral grains. 

Micromagnetic modelling, which explores the magnetic properties of a given grain (or a population of them) as a function of their size, shape, composition and temperature \cite{Williams-1989, OConbhui-2018}, is an important tool that can bring a far greater understanding of the magnetic recording process. Micromagnetic calculations have revealed, for example, i) that PSD structures (also referred to as vortex states due to their magnetisation being dominated by single or multiple vorticity centres), which is probably the most common type of magnetic domain structure in natural iron oxides \cite{Lascu-2018}, often are capable of retaining their remanence for timescales longer than the age of the Solar System \cite{Nagy-2017}; and ii) the theoretical discovery of a non-stable magnetic configuration in grain sizes beyond the SD threshold \cite{Wang-2022}, which can produce unexpected magnetic properties (e.g, the reduction of coercivities and thermal stability) and has the potential to greatly affect palaeointensity data \cite{Nagy-2022}.

Particle size is a very important controller of domain state and, consequently, of magnetic stability \cite{Dunlop-1997}, but geometry is also a critical factor. Although micromagnetic modelling of irregular particle morphologies through the discretization of three-dimensional finite element meshes is possible \cite{OConbhui-2018}, most of the theoretical knowledge built on micromagnetics of iron oxides is based on regular euhedral crystalline habits \cite{Schabes-1988, Williams-1989, Muxworthy-2003, Nagy-2017, Nagy-2019, Nagy-2022, Wang-2022}. Although natural processes can trigger regular crystal growth/alteration of iron oxides (such as biosynthesis of magnetite by magnetotatic bacteria \cite{Kopp-2008, Chang_2012}), tomographic techniques of iron oxides bearing rocks very often reveal irregularly shaped grains \cite{deGroot-2018, Lascu-2018, Nikolaisen-2020}. Consequently, it is important to understand how these irregular morphologies of particles might affect their magnetic properties, especially for grains within the palaeomagnetically significant PSD threshold.

\begin{figure}[ht]
\centering
\includegraphics[width=\linewidth]{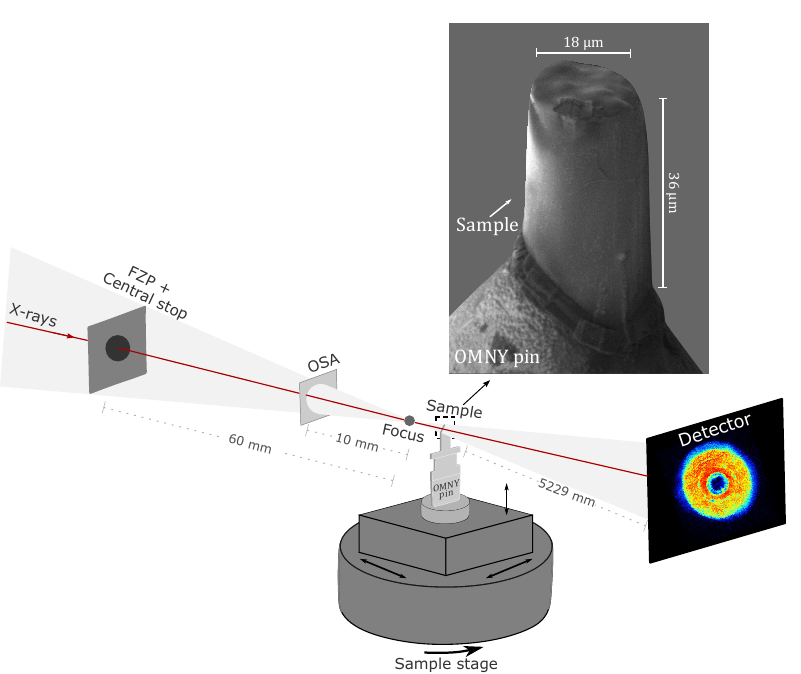}
\caption{Experimental setup for the acquisition of PXCT data. X-rays are focused by a Fresnel Zone Plate (FZP), in combination with a central stop and an order sorting aperture (OSA), to define the illumination. The radiation reaches the microscopic pillar sample, mounted at the tip of an OMNY (tOMography Nano crYo) pin \cite{Holler-2017}. A scanning electron microscopy (SEM) image of the sample is shown in the upper corner of the figure. The sample stage allows a three-axial movement of the sample on top of a rotation to perform raster scans at different angles with respect to the X-ray beam. Diffraction patterns at each scan position are captured by a pixelated detector. The components sketched in this figure are not to scale. Details can be found in the text (or in the Supplementary File).}
\label{fig: methodology}
\end{figure}

In this study, we introduce the use of Synchrotron-based Ptychographic X-ray Computed Nano-tomography (PXCT) \cite{Dierolf-2010} (Figure \ref{fig: methodology}) to image the magnetic minerals within a microscopic sample of a Neoproterozoic remagnetised carbonate, and further model their micromagnetic behaviour through finite element numerical models. The primary remanence (acquired during pre-to-syn diagenetic conditions) of carbonate rocks is very often substituted by a secondary chemical remanence (CRM) \cite{Jackson-2013, Elmore-2012} when new magnetic minerals grow or are chemically altered below their respective Curie temperature (Tc) \cite{Dunlop-1997}. These processes result in magnetic grains with a wide range of sizes and shapes, most of them within the SD/PSD grain size range, making them a target for micromagnetic studies of individual particles at the micro/nanoscopic scale. To image them, PXCT was chosen for being a non-destructive coherent diffractive imaging method that allows a quantitative mapping of the tridimensional electron density distribution of the specimen at unmatched nanoscale resolution \cite{Holler-2014,michelson2022}. Our results provide new insights into the magnetic properties of natural remanence carriers in rocks and their thermomagnetic stability.

\section*{Results}

\subsection*{Nano-tomography}

The processed tomograms representing the 3D electron density distribution of the specimen resulted in artefact-free volumes with a voxel size of $34.84 \ \text{nm}$. An example is shown in Supplementary File, Figure S1a, while details about the PXCT measurements and data processing can be found in the Methods section and in the Supplementary File. By using the accumulated histogram distribution of all of the data and the natural breaks in the distribution (Supplementary File, Figure S1c), we have segmented five compositional classes (Supplementary File, Figure S1d, Table S1). Our interpretations of the imaged materials presented here are also supported by previous chemical/magnetic analyses performed by Bellon et al. \cite{Bellon-2023a}. 

\begin{figure}[ht]
\centering
\includegraphics[width=\linewidth]{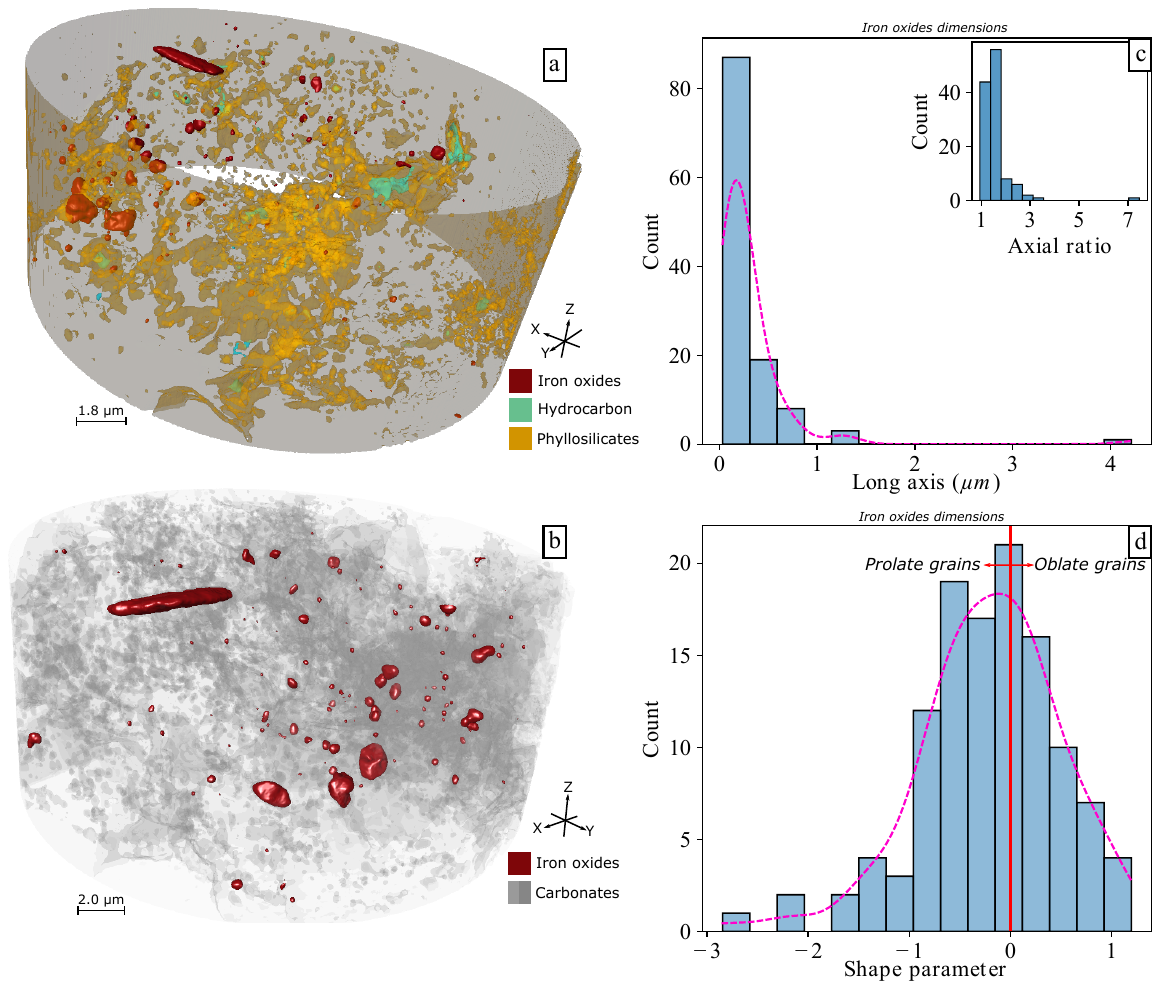}
\caption{Reconstruction of the PXCT 3D electron density distribution after segmentation, highlighting different materials in (a) and (b). Most of the iron oxide grains have a major elongation axis between $80-600 \ \text{nm}$ (c) and an axial ratio ($\text{P}_1/\text{P}_3$) $< 2$. In (d), the shape parameter ($\epsilon$, revisit Equation \ref{eq:Shape parameter}) is quite scattered, showing a well-distributed population of triaxial ($\epsilon \approx 0$), prolate ($\epsilon << 0$) and oblate ($\epsilon >> 0$) grains. }
\label{reconstruction}
\end{figure}

 The classes are named according to their increasing electron density (Figure \ref{reconstruction}). Class 1 yields the lowest electron density of $\mu = 0.31 \pm 0.11 \ e^{-} \cdot \text{Å}^{-3}$, also showing restricted volumetric occurrence. This material appears in small clusters within the sample (Figure \ref{reconstruction}a, green colour), also showing a spatial correlation to localised porosities. We attribute this class to organic matter, specifically to bitumen. The geological formation (from which our sample comes) is naturally enriched in hydrocarbons \cite{Font-2006}, as biodegradation of the organic matter is also considered partially responsible for the growth of authigenic ferromagnetic minerals. Natural bitumen density might vary depending on its formation and preservation, but the density of natural bitumens \cite{Lapidus-2018} has been reported to be $1.02-1.03 \ \text{g/cm}^3$, which reflects an electron density (Equation \ref{eq:Electron density}) of $0.30-0.31 \ e^{-} \cdot \text{Å}^{-3}$.  
 
 Class 2 $(\mu =0.65 \pm 0.08 \ e^{-} \cdot \text{Å}^{-3})$ material occupies a large volume of the sample (Figure \ref{reconstruction}a, yellowish colour), and it is interpreted as a phyllosilicate. Smectite-illite has been previously identified as a major mineralogical constituent of these samples, see Ref. \cite{Bellon-2023a}, as their conversion process (smectite to illite) is believed to release iron that sequentially reacts with the available oxygen to form iron oxides \cite{Hirt-1993,Woods-2002}. The density of such phyllosilicates will depend on their purity. Smectite's (end member smectite, \emph{EMS}, $< 2.3 \ \text{g/cm}^3$) transformation to illite might be partial (illite-rich mixed-layer clays, \emph{IML}, $2.4 - 2.7 \ \text{g/cm}^3$), or complete (end-member illite, \emph{EMI}, $> 2.7 \ \text{g/cm}^3$) \cite{Totten-2002}. The spread distribution of Class 2 matches the expected density of a partial smectite-illitisation, as \emph{EMS} and \emph{IML} electron densities are between $0.69$ and $0.72 \ e^{-} \cdot \text{Å}^{-3}$ (respectively).
 
 Classes 3 $(\mu =0.80 \pm 0.01 \ e^{-} \cdot \text{Å}^{-3})$ and 4 $(\mu =0.88 \pm 0.03 \ e^{-} \cdot \text{Å}^{-3})$ comprise the majority of the rock matrix and are interpreted, essentially, as calcium carbonates (Figure \ref{reconstruction}b). Class 3 matches the expected electron density of calcite $(\text{CaCO}_3,0.81 \ e^{-} \cdot \text{Å}^{-3})$, while Class 4 matches that of dolomite $(\text{CaMg(CO}_{3}\text{)}_{2},0.85 \ e^{-} \cdot \text{Å}^{-3})$. 

Class 5 gathers the highest mean electron densities of our sample $(\mu =1.08 \pm 0.09 \ e^{-} \cdot \text{Å}^{-3})$, almost 20\% larger than Class 4. The volumetric distribution of the material represented by this class is also quite distinct from the other ones, as these appear as individual grains spread in the matrix (Figure \ref{reconstruction}b). They also represent a wider electron density distribution, with values that reach as far as $1.24 \ e^{-} \cdot \text{Å}^{-3}$. We interpret this material as the iron oxide particles in our sample. The electron density upper limit of Class 5 was also observed for partially oxidised magnetite/maghemite in previous PXCT studies \cite{Maldanis-2020}, but is slightly lower than expected from their theoretical densities. Several aspects could implicate such discrepancies, for example:  (i) chemical impurities, crystalline imperfections and oxidation; (ii) density is size-dependent in nanomaterials, being lower/greater than a correspondent macrometric material depending on the lattice parameter \cite{Nanda-2012}. The most likely reason though is related to the partial volume effect, \emph{i.e}, for a given small volume the measured electron density has a large contribution from the voxels at the interface with its surrounding material, which can affect the final average density value. In our case, we have particles of a few hundred nanometers sampled with a 3D resolution, therefore a large proportion of the selected volume for density quantification may have a contribution from the density of the surrounding material.  If the surrounding material has a lower density, the density will be underestimated. Partial volume effects have been considered before, e.g., in the segmentation of volumes of nanoporous material obtained by PXCT \cite{ihli2017}. Prior applications of synchrotron-based techniques, specifically X-ray Fluorescence (XRF) and X-ray Absorption Near Edge Spectroscopy (XANES), have been utilised to identify magnetite/maghemite in a sister sample currently undergoing nano-tomographic analysis \cite{Bellon-2023a}. This additional evidence further supports our interpretation, as we follow the micromagnetic modelling of their properties. 

Following the processing, segmentation, and mesh generation (refer to \emph{Methods} for clarity), we have identified 118 distinct grains of iron oxides with varied shapes and sizes. Due to their high irregularity, we will discuss their sizes based on the ratios of the grains' main axes ($\text{P}_1 > \text{P}_2$ > $\text{P}_3$). More than 95\% of the major axis have an elongation of $80-600 \ \text{nm}$ (see Figure \ref{reconstruction}c), and 68\% is smaller than $250 \ \text{nm}$. The axial ratio $(\text{P}_1/\text{P}_3)$ evidences that although these grains are irregular, the size of major and minor axes of most of them do not show big discrepancies, which is reflected in the overall rounded morphologies (Figure \ref{reconstruction}c). The smallest particles of our sample are within the spatial resolution of the tomography and might carry uncertainties in the imaged morphology. Regardless, these morphologies still yield scientific validity, as any irregular shapes are valid for our micromagnetic models. We can more effectively access the morphologies by evaluating all three axes through the shape parameter ($\epsilon$ as per Equation \ref{eq:Shape parameter}). The great majority of them are triaxial to oblate-like particles ($\epsilon \geq 0$), but there is still a considerable amount of prolate-like grains ($\epsilon < 0$, Figure \ref{reconstruction}d). The largest particle identified in our sample, quite isolated in size, is a very oblate ($\epsilon > 0$) grain, which also corresponds to the biggest axial ratio in Figure \ref{reconstruction}c at a value just over 7.0. The presence of larger grains poses a computational challenge for micromagnetic modelling. Generating finite element meshes for their morphology while adhering to the exchange length ($\text{I}_\text{e}$) \cite{Rave-1998} as the maximum element size leads to excessive processing times. Consequently, we have focused exclusively on the finest grains ($\le 1000 \ \text{nm}$), which are more likely to exhibit the vortex state. As a result, from the initial set of 118 grains, we have excluded five particles in the subsequent micromagnetic modelling steps, working with a total of 113 grains. A spreadsheet with the morphological characteristics and summarised magnetic properties (which are further discussed) of all of these grains is also provided in a public repository \cite{Bellon-2024}. 

The magnetic grains we image in this work result from growing mechanisms that occurred below the magnetite's Curie temperature in a sedimentary rock, i.e., they originally bear a chemical remanence (CRM). In terms of micromagnetic modelling, the origin of these grains is not relevant to our calculations, as the major parameters influencing our results are the dimensions, morphology, and composition of these grains \cite{OConbhui-2018}. In our case, grains were discretised as finite element meshes of a material with a homogeneous chemical composition. Comparing morphologies is not a direct task, but the range of morphologies depicted in Figure \ref{reconstruction}d is similar to that of inclusions of magnetic minerals hosted in igneous rocks \cite{Nikolaisen-2020}. Therefore, we consider the implications of our further discussions do apply to any kind of rocks bearing irregularly shaped nanomagnetic grains.

\subsection*{Micromagnetics}

\subsubsection*{Local Energy Minimum (LEM)}

In this section, our objectives are (i) to assess the stability of remanence by investigating domain states; and (ii) we seek to explore the conditions under which different domain states are favoured at varying temperatures. Furthermore, we identify the temperature intervals during which these domain states exhibit dominance.

Identifying the magnetic structures of each grain involves running several models for each grain's geometry to search for local energy minimum (LEM) in the magnetic domain states (See \emph{Methods}) \cite{OConbhui-2018}. The LEM with the lowest energy will usually represent the most stable domain state for each grain. Model LEM solutions were obtained using MERRILL \cite{OConbhui-2018} (an open-source 3D micromagnetics software package), where the magnetic domain states structures are represented by a unit vector at each node of the model mesh geometry. The net magnetizations of grains are reported here as values normalised by their saturated magnetic state. Both the remanence (Mr) values and the topology of the magnetic domain structures observed for magnetite and maghemite were very similar (Supplementary File, Figure S2a), as were the corresponding values of their magnetic free energies (Supplementary File, Figure S2b). This is to be expected since the minerals share very similar magnetic material properties (Table \ref{tab: Magnetic_properties_table}). Thus although most of the results we describe here are for magnetite, we anticipate very similar behaviour for maghemite.

\begin{figure}[h]
\centering
\includegraphics[width=12.2cm]{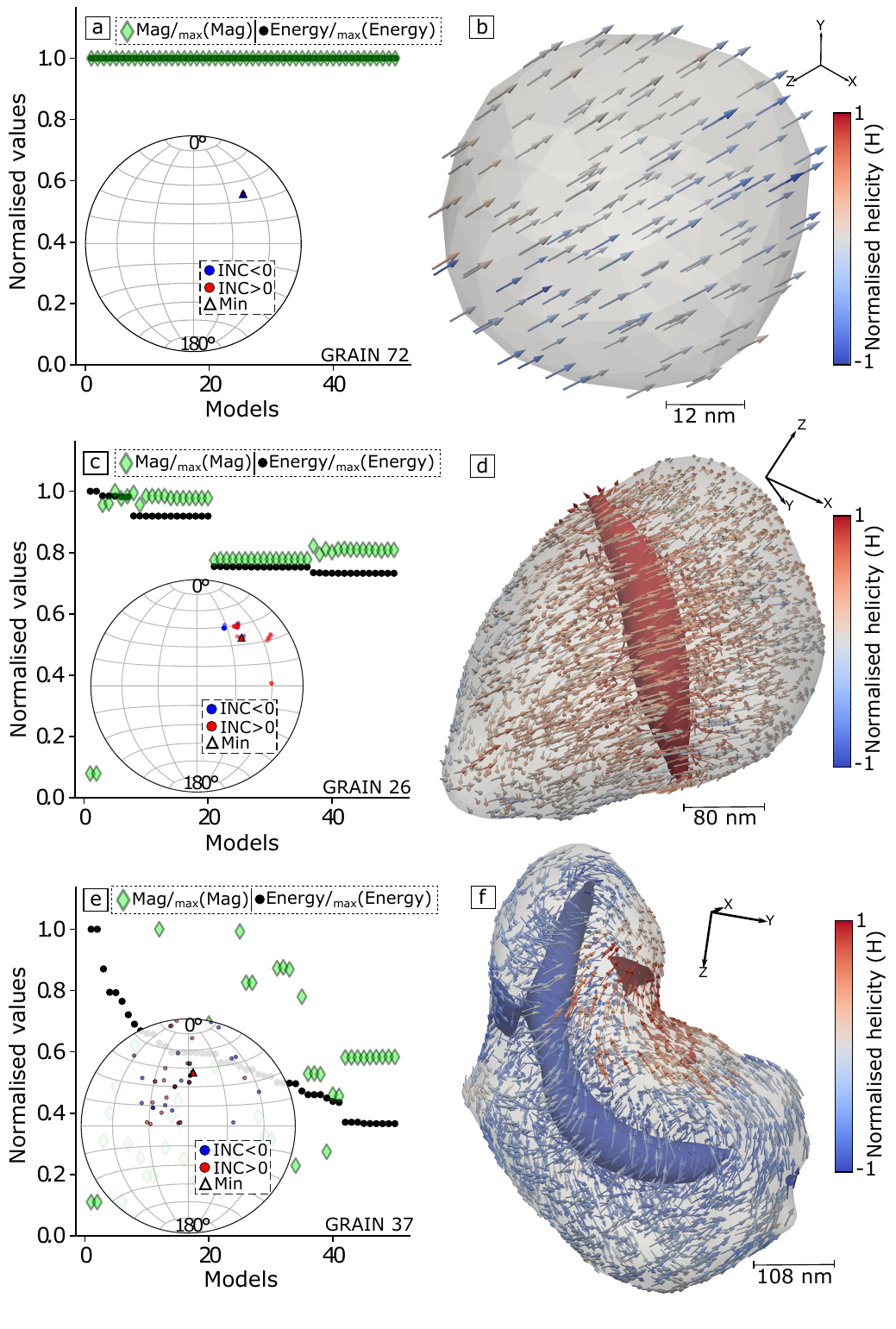}
\caption{Examples of different types of domain states (for magnetite) and their behaviours are presented here. On the left side, we have compiled the magnetisation (green diamonds) and the corresponding magnetic energy (black dots) for each of the 50 LEM models of a specific grain. These values are normalised by the respective maximum values (of either magnetisation/energy) among the fifty models. The stereographic projections within the plots display the direction of the magnetisation for all models by converting the magnetisation components in the X ($\text{M}_\text{x}$), Y ($\text{M}_\text{y}$), and Z ($\text{M}_\text{z}$) directions to declination $(\text{DEC} = \arctan (M_y/M_x))$ and inclination $(\text{INC} = \arctan(M_z/\sqrt{M_x^2 + M_y^2}))$. The inclination is coloured according to its polarity (red for positive and blue for negative), and the triangle indicates the direction of the LEM state with the lowest energy (Min). On the right, the minimum free energy model of the corresponding grain is depicted, where the arrows indicate the direction of the magnetisation, and the vortex structures inside are the isosurface hued by the helicity. SD grains show (a,b) practically no difference between the models, whereas SV can exhibit a variety of possible domain states (c,d) with distinct magnetisation (and associated energies), especially when the morphology of the grain is highly irregular (e,f).  }
\label{Types of Domain}
\end{figure}

We observe a great variety of domain states in the studied grains (Figure \ref{Types of Domain}), which we categorise following the same approach of other studies \cite{Nagy-2017,Lascu-2018,Nagy-2019,Nagy-2022}. As expected, SD states dominate the smaller particles, which in our case are also the most rounded morphologies (Figure \ref{Types of Domain}a,b). With increasing grain sizes and as morphologies become more irregular, non-uniformly magnetised domain states arise as: single-vortex (SV) structures (Figure \ref{Types of Domain}c,d), with a vortex core aligned along a specific axis; and multi-vortex (MV) structures, which only appear for the largest grain sizes considered in our study. As grain size increases it becomes possible to nucleate multiple different domain states. These different domain structures can sometimes simply be  SV states, where different LEM vortex core alignments may occur within an irregular grain morphology (Figure \ref{Types of Domain}c). Larger grains ($\text{P}_1>200 \ \text{nm}$) might result in a variety of domain states with differently twisted vortex cores, or high energy LEMS's with multiple vortex cores in the same grain (Figure \ref{Types of Domain}e).

The normalised magnetisation of magnetic particles has important implications for the identification of a given domain stature through bulk magnetic properties \cite{Dunlop-1997}, particularly for non-uniformly magnetised grains, which likely dominate the assembly of stable magnetic carriers in rocks \cite{Lascu-2018}. When we examine the range of a grain's LEM magnetisation that can be nucleated in each grain, our results indicate that most grains have $\text{Mr}<0.5$, where SV structures are dominant. None of the lowest energy states observed in the LEM models of grains with multi-vortex structures, whether composed of magnetite or maghemite, show a Mr exceeding $0.5$. Remanences greater than such threshold for MV particles have only been observed for higher energy states, which are less stable, while $\text{Mr}>0.5$ (as in Supplementary File, Figure S2a) are essentially displayed by either SD or SV-like structures. 

The temporal stability of each grain was determined using the Nudged–Elastic-Band (NEB) method \cite{Nagy-2017}. This allows us to determine the minimum energy path (MEP) between any two given domain states. For each grain, we took the lowest energy state (from a set of 50 random LEM states) and calculated the minimum energy path to its anti-parallel state. In the case where a significant new minimum was observed along the MEP, we subdivided the path at the new intermediate minimum and searched the grain for a new MEP. In this way, we obtained the energy barriers only for the lowest energy LEM state. In this study, we have chosen not to investigate the stability of higher energy (and thus less abundant) states we encountered. The energy barriers from the MEP were converted into magnetisation relaxation times $(\tau)$ using the Néel Arrhenius (Equation \ref{eq:Relaxation}).

\begin{figure}[h]
\centering
\includegraphics[width=\linewidth]{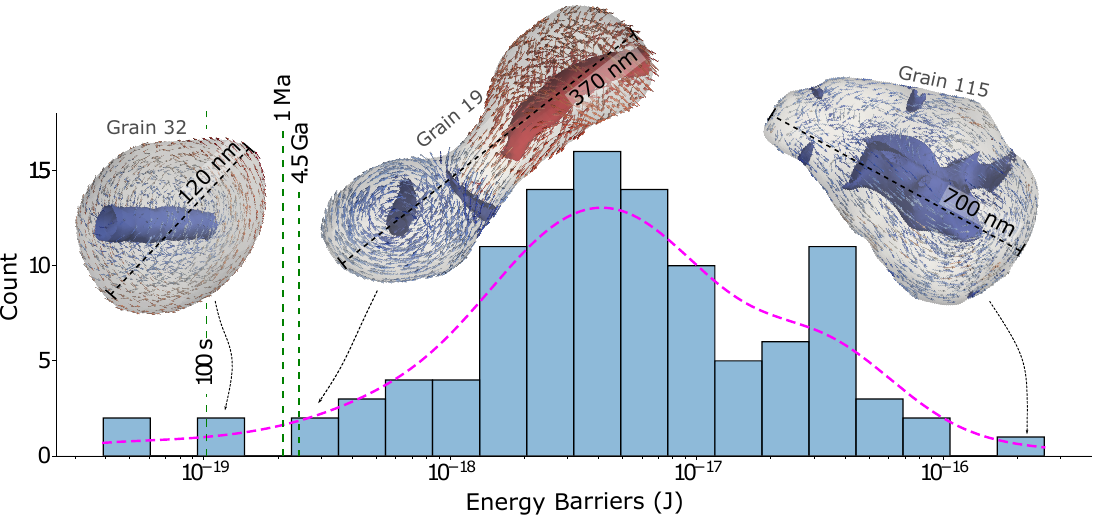}
\caption{Compilation of the computed magnetic energy barriers for each of the lowest energy LEM states across a minimum energy path to its anti-parallel state, using the Nudged–Elastic-Band \emph{(NEB)} method \cite{Nagy-2017}. This energy barrier is then applied to the Néel–Arrhenius \cite{Neel-1955} equation to calculate the relaxation time (at 20 $^ \circ C$) for the transition  $\tau$ from one domain state to another. The magenta curve represents the density of the state's distribution. The vertical dashed lines correspond to $\tau$ for $100$ seconds, $1$ million years, and $4.5$ billion years. Most of the states exhibit stabilities that far surpass the age of the Solar System (as seen in the MV states for Grains 19 and 115). However, a minority of them, such as Grain 32, are unstable magnetic recorders, representing the magnetically unstable zone (MUZ) between the transition of the SD to the ESV state. The colours shown for the arrows within the grains (as well as the vortex structures) have the same meaning as the examples shown in Figure \ref{Types of Domain} (see colour bar for helicity). Calculations were performed assuming magnetite as the mineral phase.  }
\label{NEB}
\end{figure}

In Figure \ref{NEB} we show a spectrum of energy barriers, where the vast majority of grains yield relaxation times that far surpass the age of the Solar System. The "eternal" stability includes SD, SV and MV states, the latter having the highest energy barriers. However, four grains have small $\tau$ values. Grain 32 ($\epsilon = 0.78, \ \text{P}_1= 122 \ \text{nm},  \ \text{P}_1/\text{P}_3= 1.27$), for example, yields a single vortex configuration (Figure \ref{NEB}), but our simulation indicates that thermal energy would overcome magnetic energy in $\tau \approx 3 \ \text{hours}$. The other three grains are also elongated but of grain sizes close to our tomography pixel size (e.g., Grain 113, $\epsilon = -0.35, \ \text{P}_1= 49 \ \text{nm}, \ \text{P}_1/\text{P}_3= 1.21$). Their relaxation times are mostly instantaneous (Figure \ref{NEB}), which points out that not only size but morphology, is critical in determining domain stability for such small grain sizes.

\begin{figure}[h]
\centering
\includegraphics[width=14cm]{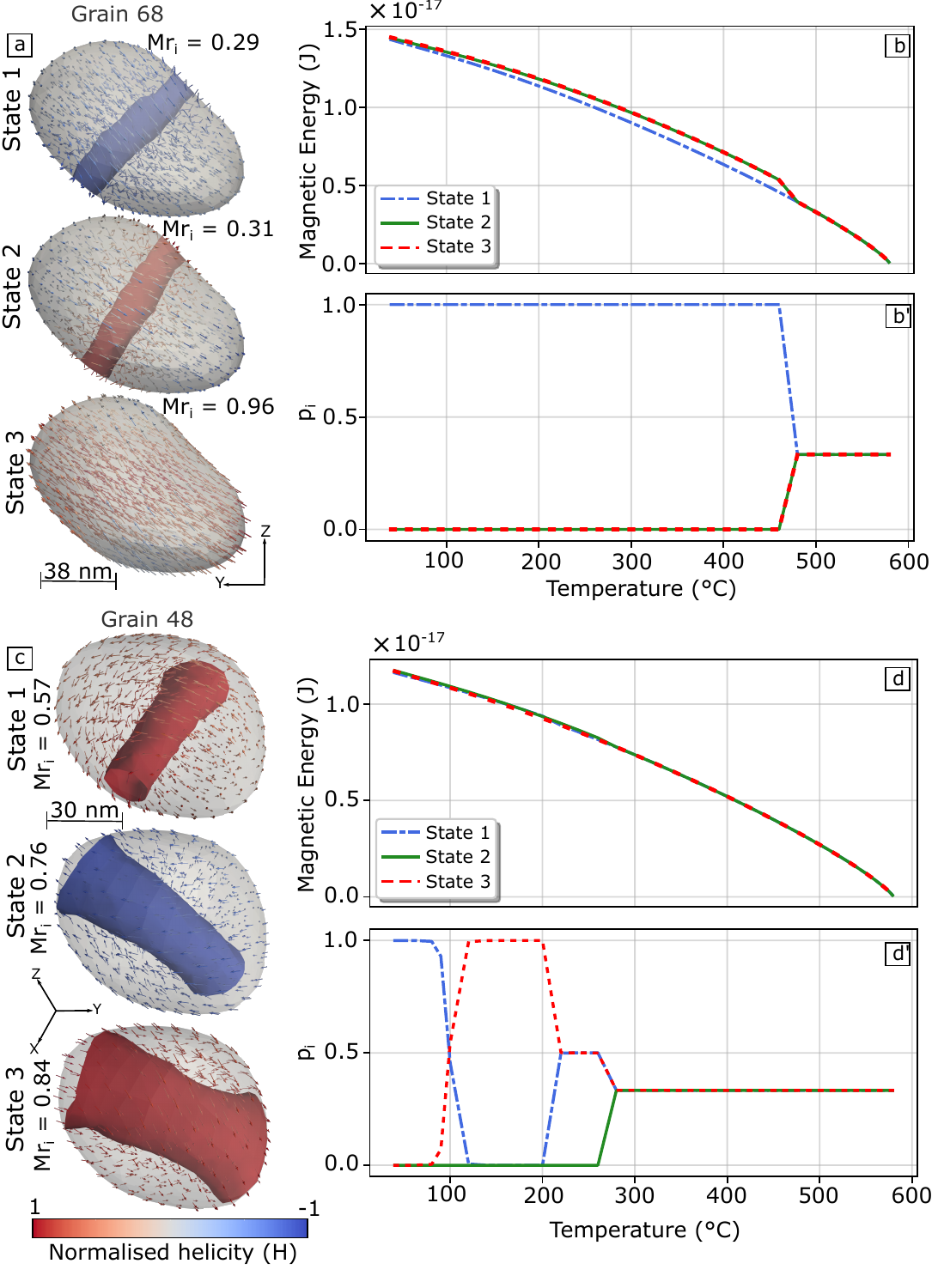}
\caption{Multiplicity of domain states for grains of magnetite. States with similar energies but distinct magnitudes of magnetisation and their temperature dependence are explored. In the first example (a), States 2 and 3 exhibit slightly higher energy with a temperature increase (b). The Boltzmann probability ($p_i$, Equation \ref{eq:Boltzmann_probability}) demonstrates that State 1 will prevail over most of the evaluated temperature range (b'). However, after reaching a certain temperature threshold (approximately $450\ ^ \circ \text{C}$), the probabilities of these domain states to occur become equal. In the example in (c), this effect is more problematic, as the energies of the states remain practically the same throughout the entire range of temperatures (d), leading to swapping of probabilities between the states (d') up to nearly $300\ ^ \circ \text{C}$, from which the $p_i$ becomes equally distributed. }
\label{domain_probability}
\end{figure}

As stated, when a particle assumes a multiplicity of possible domain states, the issue regarding which of the LEM states is the more likely to be assumed can be resolved in the presence of an isolated energy minimum, as observed in Figure \ref{Types of Domain}c,e. However, some grains may exhibit a more complex behaviour, where the LEM states yield distinct Mr but similar energies. Boltzmann probability distribution of states (Equation \ref{eq:Boltzmann_probability}), shows that for a small number of states at a given temperature, even a small difference of $2\%$ in magnetic energy (e.g., Grain 25 in Supplementary File, Figure S3) between the lowest energy LEM state and other states will result in a  $\gg 99.99\%$ occupancy rate of lowest energy state. However, the prevalence of a given state is sometimes also a function of temperature, with the most favoured domain state changing on heating or cooling.  We discuss below the case of SV/SD grains with LEM states of similar energies and distinct magnetisations. We intend to investigate how the magnetic energy of these states evolves with an increase in thermal energy, up to the Curie temperature of magnetite. Our observations indicate that the balance between magnetic and thermal energy will also depend upon both the grains' size and morphology. The SV behaviour can be classified into three groups. The first corresponds to the well-behaved type of grain, where one domain state will prevail up to the Curie temperature (Supplementary File, Figure S3). An example is Grain 25 ($\epsilon = 0.43, \ \text{P}_1= 270 \ \text{nm}, \ \text{P}_1/\text{P}_3= 3.03$), where $p_i$ indicates that a flowering state (of higher magnetisation and lower energy) will prevail over the other competing SV states. The second group represents grains where the energy of one state dominates at lower temperatures, but the magnetic energy becomes similar to other LEM states as the temperature increases (Figure \ref{domain_probability}a,b). For Grain 68 ($\epsilon = 0.60, \ \text{P}_1= 124 \ \text{nm}, \ \text{P}_1/\text{P}_3= 1.53$), e.g., a single vortex state (Figure \ref{domain_probability}b') dominates until $450 \ ^ \circ \text{C}$ but thereafter $p_i$ becomes equally distributed between the other domain states, indicating that any of these states with distinct Mr could be assumed. Finally, the third group behaves poorly. In these grains, the magnetic energies of different domain states are almost indistinguishable with increasing temperature (Figure \ref{domain_probability}c,d). For grain 48 ($\epsilon = 0.07, \ \text{P}_1= 91 \ \text{nm}, \ \text{P}_1/\text{P}_3= 1.17$), up to $\sim 300\ ^ \circ \text{C}$ there is a frequent redistribution of domain states  (Figure \ref{domain_probability}d'), and for T above $300\ ^ \circ \text{C}$ $p_i$ is equally occupied by the three different states.

\subsubsection*{Magnetic field-dependent simulations}

\begin{figure}[h]
\centering
\includegraphics[width=\linewidth]{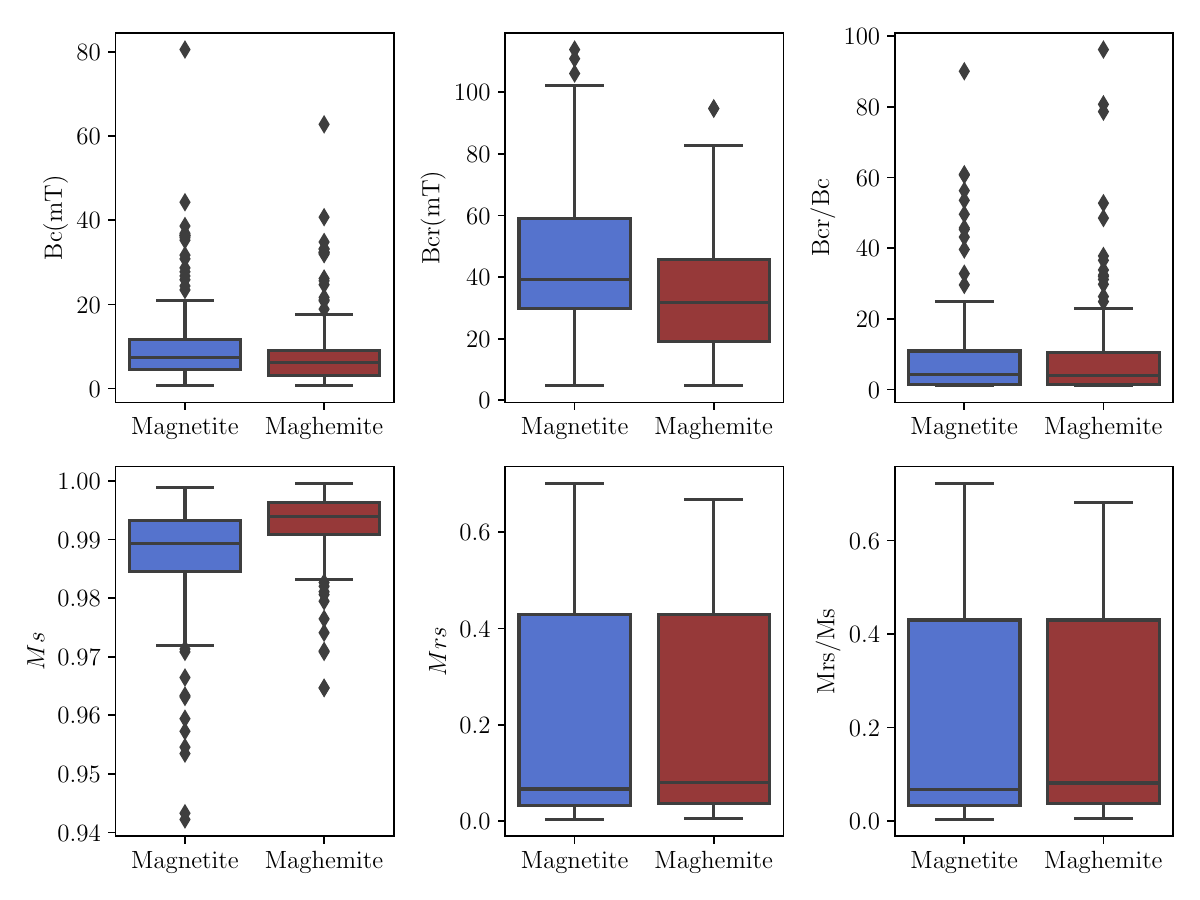}
\caption{A compilation of parameters obtained from the magnetic hysteresis loops and backfield curves for magnetite (blue) and maghemite (brown) is presented using boxplot statistics. These parameters correspond to averaged curves obtained for each grain in twenty different directions (refer to Methods for details). Bc represents the bulk coercivity, Bcr is the coercivity of remanence, Ms indicates the magnetisation saturation of the averaged loop at the maximum field, and Mrs is the remanence of saturation. Black diamonds signify statistical outliers. As illustrated, magnetite and maghemite exhibit practically indistinguishable properties. }
\label{Properties_magn_maghe_hys_back}
\end{figure}

\begin{figure}[h]
\centering
\includegraphics[width=\linewidth]{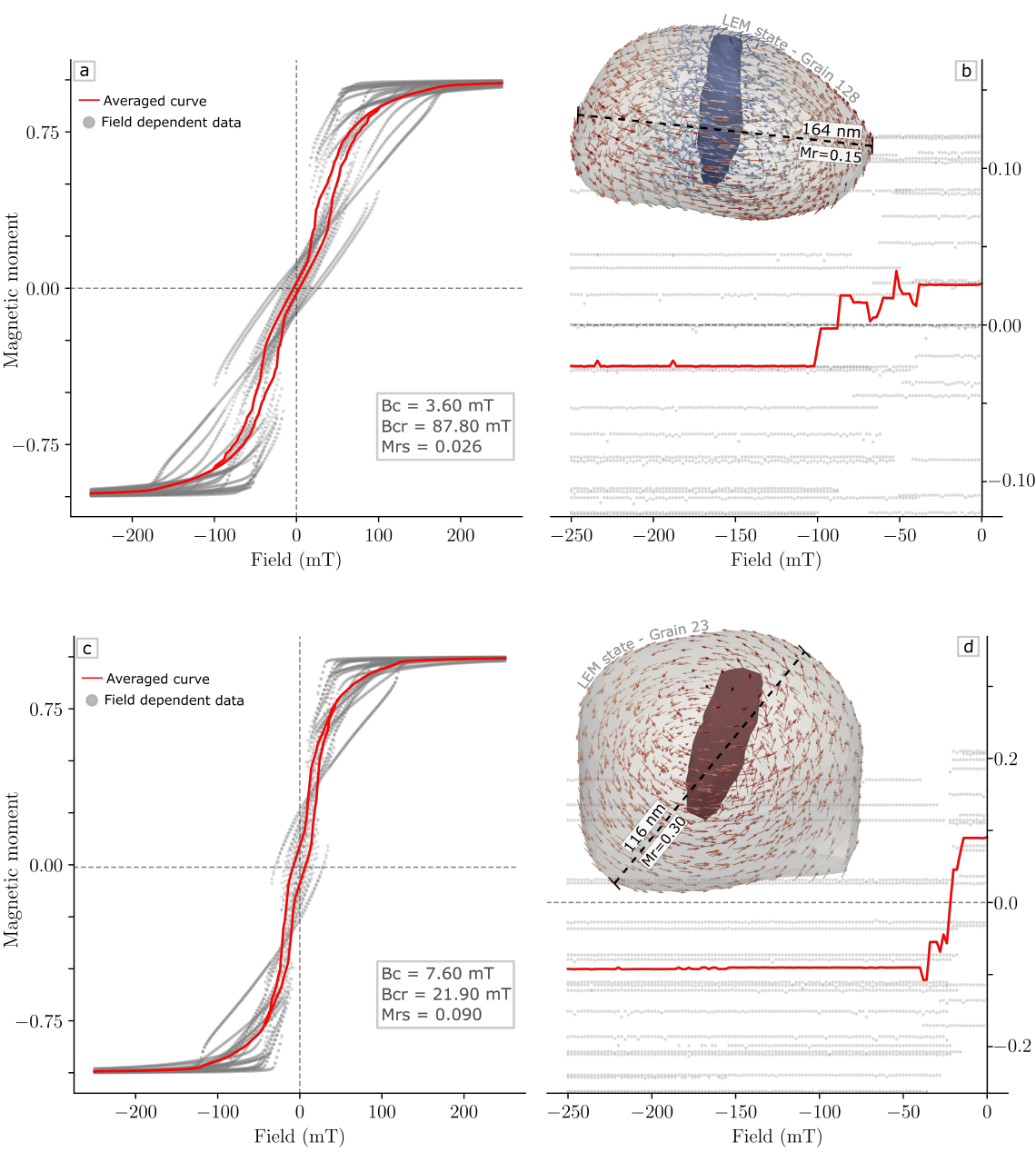}
\caption{Examples of magnetic hysteresis (a, c), backfield curves (b, d), and their parameters (for magnetite). The grey dots represent the field-dependent data, and the red curves depict their averages. Notably, for both types of simulated experiments, different field directions result in distinct curves as a result of the grain's shape anisotropy. The particles situated at the upper side of the backfield curves (b, d) correspond to the LEM states obtained at zero field. Here, we highlight how the Mr of these states might differ significantly from the Mrs of magnetic hysteresis (compare the Mrs values within the boxes in a and c with the respective Mr values within the grains). }
\label{Hysteresis-backfield-example}
\end{figure}

We have also modelled the hysteresis loops and backfield curves for each of the studied grains (see Methods), accounting for shape anisotropy effects by running models in different field directions and averaging their results. In terms of normalisation, the LEM grain's magnetisation (Mr) differs from those obtained from the saturation of remanence (Mrs) (see Figures \ref{Properties_magn_maghe_hys_back}, \ref{Hysteresis-backfield-example}). Both of these represent the remanence at zero-field. However, Mrs is determined from the interception of the averaged hysteresis loop, while the Mr is obtained through an energy minimisation starting from a random initial guess. An important effect of averaging over distinct field directions is that saturation (i.e. complete alignment of the magnetisation in the direction of the applied field) at a given field strength will strongly depend on the field direction with respect to the magnetic anisotropy axes of the grain. A particle's anisotropy is often controlled by its morphology, which then determines how easily a grain becomes magnetically saturated. Therefore, we consider the saturation parameter of magnetic hysteresis, Ms, as the corresponding value of magnetisation for each averaged curve at $250 \ \text{mT}$. We apply boxplot statistics to discuss some of the magnetic properties in terms of all grains. As also highlighted in the previous subsection, magnetite and maghemite properties are very similar producing practically indistinguishable results (Figure \ref{Properties_magn_maghe_hys_back}). To avoid redundancies, we henceforth discuss only the properties of magnetite.

Except for a few outliers, the Ms is practically the same as the theoretical one for magnetite ($0.99 \pm 0.01 $). Mrs span from $0.0035 - 0.70$, whereas only SD and SV states occupy values greater than $0.5$. Significant discrepancies emerge when comparing Mr with Mrs. Firstly, there is an important influence of the field direction in the whole averaged loop, which can generate very different curves with distinct Mrs for each field direction (Figure \ref{Hysteresis-backfield-example}a,c). For grains whose minimum energy models yield $\text{Mr} \gtrapprox 0.5$, the resulting Mrs from the simulated random distribution of grains can be 50\% smaller  (Supplementary File, Figure S4a,b).  The difference is even greater for particles whose $\text{Mr} \lessapprox 0.5$, which can result in Mrs up to $80\%$ smaller than the magnetisation of its lowest LEM (Mr) (Figure \ref{Hysteresis-backfield-example}). This effect increases with grain size, the larger grains being able to accommodate a larger number of domain states each with a distinct remanence, but this effect is consequently averaged during the field-dependent magnetic experiments or numerical simulations of random distributions of such grains. Bulk coercivities (Bc) span over a large interval ($0.64-80.59 \ \text{mT}$, Figure \ref{Properties_magn_maghe_hys_back}), most of them within expected \cite{Tauxe-2010} values for magnetite ($11.41 \pm 11.86 \ \text{mT}$). The coercivity of remanence (Bcr) obtained from the backfield curves is also within the expected range ($46.27 \pm 25.32 \ \text{mT}$), but several grains present higher values (e.g., Figure  \ref{Hysteresis-backfield-example}b), some of these being greater than $100 \ \text{mT}$. 

\begin{figure}[h]
\centering
\includegraphics[width=\linewidth]{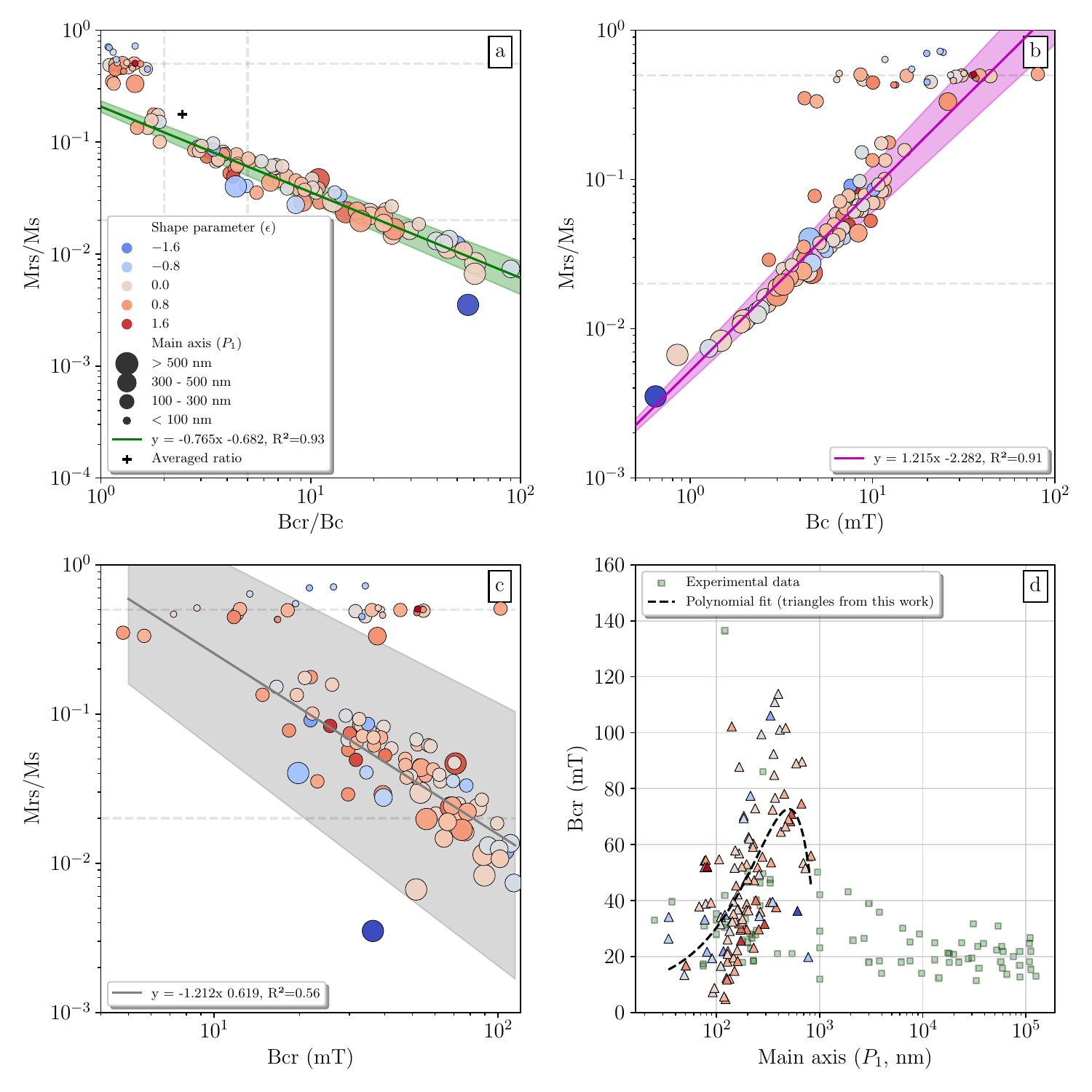}
\caption{a) Remanence (Mrs/Ms) and coercivity (Bcr/Bc) ratios are compared in the Bilogarithmic Day plot \cite{Day-1977, Dunlop-2002}. Two distinct clusters are evident: the first, located near $\text{Mrs/Ms} \approx 0.5$, and the second, following a linear trend (green line with shaded extension representing $\alpha 95$). For a, b and c, the circles are colour-coded based on the shape parameter ($\epsilon$), with blue/red indicating more prolate/oblate grains. The circle size represents grain size, and the black cross (in a) indicates the average ratio of all modelled grains. Neél's diagram \cite{Neel-1955} (b) and its sister plot (c) depict linear tendencies with confidence areas. Similar clusters are observed in both plots, indicating an increase in remanence with coercivity (Bc) and a decrease with the increase of coercivity of remanence (Bcr). In plot d, Bcr values obtained in our models (triangles) are compared with their respective grain sizes, while green squares represent experimental data \cite{Day-1977,Levi-1978,Dankers-1981,Ozdemir-1982,Bailey-1983,Dunlop-1986,Argyle-1990,Heider-1996,Muxworthy-2000,Muxworthy-2006,Krasa-2011,Almeida-2015}. A third-order polynomial fit serves as a guide, revealing a trend where Bcr peaks around $600-700 nm$ before decreasing, consistent with experimental findings.}
\label{Dayplot_magnetite}
\end{figure}

The remanence (Mrs/Ms) and the coercivity (Bcr/Bc) ratios are important indicators of magnetic domain states \cite{Dunlop-2002}. In Figure \ref{Dayplot_magnetite}, we further explore these ratios by making use of the bilogarithmic Day plot (Bcr/Bc vs Mrs/Ms) \cite{Day-1977,Dunlop-2002}, as well as the Néel's diagram (Mrs/Ms vs Bc) \cite{Neel-1955} and its sister plot (Mrs/Ms vs Bcr). The Day plot carries markings of conventional divisions  \cite{Day-1977,Dunlop-2002} for SD ($\text{Mrs/Ms} \geq 0.5, \ \text{Bcr/Bc} \leq 2 $), PSD ($ 0.5 >  \text{Mrs/Ms} > 2x10^{-2} , \ 5 >  \text{Bcr/Bc} > 2$) and MD ($ \text{Mrs/Ms} > 2x10^{-2}, \  \text{Bcr/Bc} > 5$) states (Figure \ref{Dayplot_magnetite}a). These limits, as well as the interpretation of such parameters, have been questioned \cite{Roberts-2018,Lascu-2018} to be non-representative of the many processes that can lead to variations of such ratios. Regardless, we keep those fields simply as guides to the eye. On the Day plot, there are two well-defined clusters: for the first one, SD grains and some SV grains group near $\text{Mrs/Ms}=0.5$; for the second one, grains follow a linear trend, $\text{y}=(-0.765 \cdot x) - 0.682, \ \text{R}^{2}=0.93$. The smallest grains ($100 \ \text{nm} \leq $) are within the first cluster (Figure \ref{Dayplot_magnetite}a). The grains aligned along the linear trend, which includes both SV and MV states, extend from the original PSD down to the MD region. Besides such an obvious mismatch with the traditional PSD field, some SV states exhibit extremely high Bcr/Bc ratios (Figure \ref{Dayplot_magnetite}a), especially associated with larger grains. When all of the backfield and hysteresis are averaged together, our sample produces averaged ratios of $\text{Mrs/Ms}=0.117$ and $\text{Bcr/Bc}=2.45$, which falls within the original PSD field of the Day plot (Figure \ref{Dayplot_magnetite}a).

On Néel's diagram (Figure \ref{Dayplot_magnetite}b), the same clusters are also present, which is expected since they share the same ordinate axis. Bc of the first cluster has a quite spread behaviour, but the other grains show a linear increase of remanence with the bulk coercivity, $\text{y}=(1.215 \cdot x) - 2.282, \ \text{R}^{2}=0.91$. This relationship is inversely proportional to the coercivity of remanence (Figure \ref{Dayplot_magnetite}c) for which, although more dispersed, Bcr increases as the remanence ratio decreases, $\text{y}=(-1.212 \cdot x) + 0.619, \ \text{R}^{2}=0.56$. For all the properties here discussed, there is no clear relationship between a magnetic property and the respective elongation parameter of a grain. However, even for these irregularly shaped grains, there seems to be a correlation between the magnitude of Bcr and the grain size. When we put together experimental data \cite{Day-1977,Levi-1978,Dankers-1981,Ozdemir-1982,Bailey-1983,Dunlop-1986,Argyle-1990,Heider-1996,Muxworthy-2000,Muxworthy-2006,Krasa-2011,Almeida-2015} and the results of our models, despite a large spreading, the coercivity of remanence seems to increase and peak around grain sizes between $600-700 \ \text{nm}$, from which it tends to decay again (Figure \ref{Dayplot_magnetite}d).

\section*{Discussion}

\subsection*{On the magnetic properties of natural nanoscopic grains}

Although not surprising, an important aspect of our results is that the magnetic properties of magnetite and maghemite, as demonstrated in Figure \ref{Properties_magn_maghe_hys_back}, as well as their domain states, are practically indistinguishable at ambient temperatures. The challenge of differentiating between these two magnetically soft minerals can be addressed through the use of temperature-dependent experiments \cite{Tauxe-2010}, which are proven to be essential for properly interpreting environmental magnetism data and inferring exogenous processes \cite{Qingsong-2012}, but often expose the samples to chemical or domain state change. 

Mrs/Ms and Bcr/Bc ratios derived from magnetic hysteresis and backfield curves can be characteristic for an entire assemblage of grains. It is natural, therefore, that various processes impact the comparison of these ratios in the Day plot \cite{Day-1977, Dunlop-2002}. Such processes include mixtures of magnetic minerals with distinct coercivities, and/or mixtures of grains in distinct domain states \cite{Tauxe-1996, Bellon-2023b}, as well as surface oxidation, mineral stoichiometry, and the intrinsic magnetic properties of each grain \cite{Roberts-2018}. Nevertheless, our data demonstrates that for both magnetite and maghemite (Figure \ref{Dayplot_magnetite}a), two well-defined clusters exist. One is occupied by higher remanences (solely featuring SD and SV states), and the other displays a linear trend cutting across the Day plot towards lower remanence ratios and higher coercivity ratios. The Day plot trend we observe for SV grains closely follows that found in idealised grains by Williams et al \cite{Williams-2024}, although, for our irregular grains, we do not observe a well-defined correlation with grain shape. As grain sizes increase, so does the internal demagnetising field \cite{Williams-1989, Dunlop-1997}, until it grows sufficiently strong to collapse into a lower energy domain state. Lower energy domain states often imply a decrease in remanence, which we observe in our analysis with the increase in Bcr (Figure \ref{Dayplot_magnetite}b). Our data also suggests a peak for Bcr with the increase of grain size around $600-700 \ \text{nm}$  (Figure \ref{Dayplot_magnetite}d), aligning considerably with experimental data of grains with distinct morphologies and sizes \cite{Day-1977,Levi-1978,Dankers-1981,Ozdemir-1982,Bailey-1983,Dunlop-1986,Argyle-1990,Heider-1996,Muxworthy-2000,Muxworthy-2006,Krasa-2011,Almeida-2015}. Generally, our results also align well with data derived from iron oxide inclusions \cite{Nikolaisen-2020}.

In our simulations, the influence of superparamagnetic (SP) grains is not resolved. The combination of all grains in hysteresis and backfield curves suggests that a high quantity of SP particles could potentially decrease Mrs/Ms and increase Bcr/Bc \cite{Dunlop-2002}. Careful interpretation of these diagrams is necessary. However, with well-established magnetic mineralogy supported by other analyses such as thermal demagnetisation of synthetic remanences, thermomagnetic curves, first-order reversal curves (FORCs), and chemical/imaging analyses, the Day plot, Néel's diagram, and related plots can provide useful insights. 

Magnetic hysteresis, including FORCs\cite{Roberts-2000, Roberts-2014, Roberts-2022}, heavily rely on the Mrs to infer field-induced states and not remanence states. One of our crucial observations is that unless the grains are SD, the true remanence of a grain will be potentially underestimated. While this may not be a significant concern for very fine particles exhibiting SD-like behaviour ($\text{Mrs} \gtrapprox 0.5$), it can become a pronounced phenomenon for larger grain sizes. Experimentally, magnetic hysteresis, backfield curves, isothermal remanence curves (\emph{IRM}s), and FORCs are performed using a fixed field direction. As illustrated in Figure \ref{Hysteresis-backfield-example}, the applied field's direction can induce variable magnetic properties due to the grains' shape anisotropy. For a population of irregular grains randomly oriented, changing the field direction may not significantly decrease this effect because rocks might have millions of magnetic grains. However, for materials with a well-oriented magnetic fabric (e.g., inclusions in minerals, and highly deformed rocks), conducting these measurements in variable field directions (or changing sample orientation) might enhance the determination of Mrs.

\subsection*{Temporal and thermal stability of irregularly shaped grains}

The capacity of magnetic minerals to retain their remanence over geological time-scales is crucial for planetary studies, as rock-magnetic recordings provide information on Earth and Planetary evolution such as tectonics \cite{Merdith-2021}, the timing of inner core nucleation \cite{Zhou-2022a}, and even the early stages of solar nebula development \cite{Wang-2017}. Néel's theory \cite{Neel-1955} has long been employed to demonstrate that SD grains can retain their remanence over billions of years. Although this domain state is highly stable, it spans only a very limited range of sizes, and most terrestrial and extraterrestrial magnetic materials will contain a much broader range of grain sizes.

Nagy et al. \cite{Nagy-2017} have calculated the temporal stability of equidimensional truncated octahedron grains of magnetite ($80-1000 \ \text{nm}$) in the SV state and have demonstrated that, except for grains between $80-100 \ \text{nm}$, their stability surpasses the age of the Solar System. Our calculations for non-euhedral morphologies, which are more realistic for oxide grains in most natural systems, generally align with this interpretation (Figure \ref{NEB}), demonstrating that most SV and MV grains can also retain eternal stability at environmental temperatures. Nagy et al. \cite{Nagy-2017}  also pointed to a magnetically unstable zone (MUZ) for grain sizes between $80-100 \ \text{nm}$, for which the relaxation times are near instantaneous at room temperature. In the MUZ for magnetite, the SV domain state is aligned with the hard magnetisation axis (HSV), while large SV grains align their magnetisation with the easy magnetic anisotropy axis (ESV). Wang et al. \cite{Wang-2022}, have shown that within the MUZ ($68-100 \ \text{nm}$) the grains' evolution proceeds as SD $\rightarrow$ HSV $\rightarrow$ ESV. For such, grains within the MUZ were shown to occupy about $20 \ \text{nm}$ for euhedral and also for rounded particles \cite{Nagy-2019,Williams-2010}. Four of our modelled grains within the MUZ size range proposed by Nagy et al. \cite{Nagy-2017}  are unstable at room temperature, e.g. Grain 32 (Figure \ref{NEB}) with an elongation axis of $122 \ \text{nm}$. Furthermore, several of the grains (with variable morphology) that we studied have $\text{P}_1$ within the proposed range dimensions of the MUZ (e.g., Grain 135: $\epsilon = -0.93, \ \text{P}_1= 80 \ \text{nm}, \ \text{P}_1/\text{P}_3= 1.39$  | Grain 48: $\epsilon = 0.07, \ \text{P}_1= 91 \ \text{nm}, \ \text{P}_1/\text{P}_3= 1.17$), and yet they are highly stable. Therefore, we suggest that the transition of SD to HSV (consequently at the onset of the MUZ) is strongly affected by the morphology of the grain.

Figure \ref{Types of Domain} further highlights that a multiplicity of domain states may occur in one or more temperature intervals and that the likelihood of a grain being able to nucleate multiple domain states increases with grain size.   This is a particular problem for the determination of palaeointensities which are often performed using a stepwise cycle of heating, where the original remanence (a thermal remanence, TRM) in any temperature interval is replaced by laboratory-induced partial thermal remanence (pTRM) in a known field.  \cite{thellier-1959,Yu-2004}. An important assumption in these experiments is that the magnetic recorders in the sample obey the so-called Law of Reciprocity, which states that a magnetic grain demagnetises (unblocks) and re-magnetises (blocks) at the same temperature i.e. ($T_{ub} = T_{b}$) \cite{Koenigsberger-1938,thellier-1959,Tauxe-2021}, which is only valid for the original SD theory of Néel. When multiple domain states are possible in the same grain, thermal cycling may produce a pTRM with grains now occupying very different domain states with different thermal stabilities, such that $T_{ub} \neq T_{b}$. This issue has been discussed by Nagy et al. \cite{Nagy-2022}. In their models, using a euhedral magnetite grain ($100 \ \text{nm}$, with $30\%$ of elongation along [$100$]), they not only point to a multiplicity of domain states but also to a preference for a given state as temperature changes, especially as a function of the cooling ratio. A similar effect has been identified by Tauxe \cite{Tauxe-2021} where "fragile" magnetic remanences behaved differently during stepwise demagnetisation and remagnetisation after samples "aged" by sitting in various storage fields over many months. This time-induced effect is equivalent to the pure thermal effect described by Nagy et al. \cite{Nagy-2022}, which would be caused by the multiplicity of possible domain states. 

We now suggest this situation might affect a much broader range of grain sizes, and be highly dependent on grain morphology. Any failure of blocking and unblocking reciprocity, as well as the often large change in a grain's magnetisation, will make it impossible to obtain reliable paleomagnetic intensities. In exploring the multiplicity of domain states that co-exist with similar magnetic free energies for grains of irregular shape we observed this same problem. Grains whose behaviour is dominated by one or two states over a large range of temperatures are the ones that will likely provide good paleointensity results up to a certain temperature, where the difference in energy is significant. Above such temperature threshold, any of the possible domain states could be assumed as the probability is equally distributed (Figure \ref{domain_probability}b'). The exact occupancy of the different domain states will be dependent on the thermomagnetic history of the sample. Repetitive cycling through temperatures, each yielding a variety of domain states, is expected to gradually shift the overall distribution of domain states within the sample, deviating from the initial distribution that carried the original (geological) recording of the geomagnetic field. Thus, thermal cycling and differential heat/cool rates are likely to further exacerbate the problem \cite{Nagy-2022}. It is possible that grains exhibiting the two dominant state behaviours (Figure \ref{domain_probability}b') could in some instances be an alternative explanation to high temperature palaeomagnetic experimental failure that is often attributed to chemical alteration \cite{Coe-1967}. Grains with more complex behaviour (Figure \ref{domain_probability}d'), characterised by a constant re-balancing of probabilities, are expected to yield unreliable thermally dependent paleomagnetic and paleointensity data, similar to the findings presented by Nagy et al. \cite{Nagy-2022}.

Magnetic grains whose domain states are either truly SD or ESV but far from such transition zones (mostly grains $\gtrapprox 200 \ \text{nm}$) are likely to be the most stable magnetic recorders, and also the ones for which palaeomagnetic information is less affected by thermal experiments. That is probably because the mechanism for magnetic state transitioning of ESV grains is the structure coherent rotation \cite{Nagy-2017}, which (as well as for SD grains) requires a uniform rotation of all of the magnetic moments associated with the grain. Furthermore, because domain state transitions seem to imply several complications in the magnetic properties of grains, there should be an equal dedication to studying grains within the ESV $\rightarrow$ MV transition.

\section*{Methods}

\subsection*{Sample preparation}

The sample used in our research is a Neoproterozoic cap carbonate rock, part of the Guia Formation, which spreads over the Amazon craton and the Paraguay belt (central-West Brazil). These rocks have been targets of palaeomagnetic studies and suggested to carry pervasive evidence of authigenic growth of secondary magnetic grains as a result of low-temperature (\emph{<Tc}) chemical processes \cite{Font-2006,Trindade-2003}. Recently, Bellon et al. \cite{Bellon-2023a} performed a mineralogical, microscopic and magnetic analysis on remagnetized carbonate rocks from South America. We have chosen a thin-section sample from their data set that has shown remarkable amounts of micrometric/nanometric iron oxides to build continuous work. Preparation of microscopic pillars was acquired by firstly coating the thin sections with carbon and sequentially cutting a $\approx 18 \times 40 \mu \text{m}$ pillar using a FIBSEM Helios 600i TFS FIB-SEM (Scientific Center for Optical and Electron Microscopy - ScopeM, EHT Zürich). Because the pillar sample represents only an infinitesimal part of the macroscopic thin section, the pillar was prepared in a region that had been chemically mapped for iron oxides through both SEM-EDS and synchrotron-based X-ray Fluorescence/X-ray Absorption Near Edge Spectroscopy (XRF/XANES), \cite{Bellon-2023a}. Lastly, the pillar was mounted on a PXCT OMNY-pin \cite{Holler-2017} using a nanomanipulator.

\subsection*{Synchrotron-based Ptychographic X-ray Computed Nano-tomography (PXCT)}

PXCT was performed at the \emph{cSAXS} (coherent small-angle X-ray scattering) beamline of the Swiss Light Source at the Paul Scherrer Institute (PSI, Villigen, Switzerland). For the acquisition, we used the flOMNI setup \cite{Holler-2012,Holler-2014}, a photon energy of 6.2 keV, and we placed the sample downstream of the focal spot defined by a Fresnel zone plate (FZP). This energy was chosen due to the major mineral constituent being calcite $(\text{CaCO}_3)$, which results in nearly $50\%$ of transmission in the middle of the sample for a pillar of 20 $\mu$m. Scans were performed in a Fermat spiral pattern\cite{huang2014} with a field of view of $37 \times 10\, \mu \text{m}$, and coherent diffraction patterns were collected with an Eiger $1.5 \text{M}$ detector \cite{dinapoli2011} positioned $5.229 \text{m}$ downstream of the sample. We acquired $650$ projections at different incidence angles of the beam from the sample, ranging from $0$ to $ 180 ^ \circ$. During the data collection, we estimate that a dose of $5.6 \times 10^{8} \ \text{Gy}$ was delivered to the sample. Reconstruction of ptychographic projections were performed with the PtychoShelves package developed by the Coherent X-ray Scattering group at PSI \cite{wakonig2020}, using a combination of the difference map algorithm \cite{Thibault-2009}, followed by the maximum likelihood algorithm \cite{Thibault-2012}. This resulted in images with a pixel size of $34.8 \ \text{nm}$. Phase images were treated with the removal of offset and linear terms and were further aligned in the vertical and horizontal directions with a subpixel precision method \cite{GuizarSicairos-2011,odstrcil2019_alignment}. Tomographic reconstructions were finally performed by applying a filtered back projection \cite{GuizarSicairos-2011,odstrcil2019_alignment}. Using Fourier shell correlation \emph{(FSC)} \cite{vanHeel-2005}, we estimated that the average 3D half-pitch resolution of the whole 3D dataset is $89 \text{nm}$. This resolution reflects the specimen's overall average. The limited quantity of high-density magnetite particles could be imaged with better resolution due to their contrasting density with the surrounding material.  More details about the PXCT experimental setup, data collection and reconstruction can be found in the Supplementary File.

\subsection*{Segmentation and processing of tomographic data}

Distinguishing from traditional X-ray microscopy techniques, X-ray ptychography provides images of the complex-valued transmissitivity of the specimen, where the real part is related to the absorption and the imaginary part with the phase shift of the X-rays. In combination with tomography, PXCT provides the 3D distribution of the full complex index of refraction $n(\vec{r})$ \cite{Dierolf-2010}:

\begin{equation}\label{eq:full refraction index}
n(\vec{r}) = 1 - \delta (\vec{r}) - i \cdot \beta (\vec{r})
\end{equation}

where $\delta (\vec{r})$ and $\beta (\vec{r})$ are obtained from the tomographic reconstruction using the phase and the absorption images, respectively. The phase shift of X-rays is related to refraction, which describes the change in the direction of the X-rays, it is strongly associated with the material density, composition and internal structure, while absorption of X-rays is related to the local structure and composition environment of an adsorbing atom, apart from its density. In our paper, we rely on the $\delta (\vec{r})$ reconstructed tomography to recover the morphology and segment different mineral constituents of the pillar due to its higher resolution compared to the $\beta (\vec{r})$ tomogram. Nevertheless, we compare that data with $\beta (\vec{r})$ reconstruction to refine our interpretation. The quantitative $\delta (\vec{r})$ values obtained at each voxel are converted to electron density $n_{e}(\vec{r})$ as \cite{Als-Nielsen-2011}:

\begin{equation}\label{eq:Electron density}
n_{e}(\vec{r}) = \frac{\delta (\vec{r}) k^{2}}{r_{0} 2 \pi}
\end{equation}

where $k = (2 \pi / \lambda)$, $\lambda$ is the X-ray wavelength and $r_{0}$ is the Thomson scattering length. In order to compare with our data, we can use the theoretical relation between the mass density of a given material $\rho(\vec{r})$ and its $n_{e}(\vec{r})$ as \cite{Diaz-2015}:

\begin{equation}\label{eq:Material density}
\rho(\vec{r}) = \frac{n_{e}(\vec{r}) \cdot A}{N_A \cdot \phi}
\end{equation}

where \emph{A} is the molar mass, $N_A$ is Avogadro's number and $\phi$ is the total number of electrons in a molecule of the material. 

After converting the $\delta$ values to electron density (Equation  \ref{eq:Electron density}), we calculated the histogram distribution of all voxels in the tomogram to segment the distinct populations composing the sample (Supplementary File, Figure S1c). Before segmenting the data, we first deal with salt-and-pepper noise (maximum/minimum intensity voxels randomly distributed throughout the image \cite{Boncelet-2009}) by applying a three-dimensional Gaussian filter to the data (assuming an isotropic $ \sigma = 1$, centred at each voxel). Data segmentation was acquired by processing the natural breaks in the histogram distribution and using them as thresholds after stripping away the air response. Based on the $n_{e}$ distribution, five classes have been defined (Supplementary File, Figure S1d). The $n_{e}$ slices were reclassified into binary files, assigning \emph{True} for the iron oxides and \emph{False} for all of the other materials. Binary files were then imported to the software \emph{Dragonfly} to generate stereolithographic data (STL file) of each iron oxide grain. Even aware that filtering would be able to deal with salt-and-pepper noise, we have removed particles represented by a single voxel, to give more robustness to our models. 

To systematically acquire information on the morphology of the iron oxides we used a Principal Component Analysis (PCA) of their spatial distribution. Firstly, we export the STL files as a Cartesian coordinate cloud of points (x, y, and z) describing the external morphology of the particle. Sequentially, we use a Python script to calculate the convex hull \cite{scipy_spatial_convexhull} of each set of coordinates of a given particle. A convex hull is the smallest possible convex shape that encloses a given set of points, forming a polygon that connects the outermost points by ensuring that all of its internal angles are less than 180$^ \circ$ \cite{Baillo-2021}. We then performed a PCA \cite{scikit-learn}, to get the eigenvectors (main directions) and eigenvalues (their magnitudes) of a covariance matrix of standardised variables. Sequentially, we have projected the original convex hull vertices onto the space defined by the principal components and calculated the size of the hull along the main directions. In this case, the maximum variance component $\text{P}_1$ represents the maximum elongation axis, $\text{P}_2$ is the intermediary one and $\text{P}_3$ is the minimum variance component (short axis). To more broadly evaluate the morphology of the particles in our sample, we use the shape parameter (here called $\epsilon$) of Nikolaisen et al. \cite{Nikolaisen-2020} by using our main components achieved through the PCA analysis:

\begin{equation}\label{eq:Shape parameter}
\epsilon = \log\left(\frac{\text{P}_1 - \text{P}_2}{\text{P}_2 - \text{P}_3}\right)
\end{equation}

for which triaxial morphologies (all of the main axes have different dimensions) will have $\epsilon \approx 0$, the more positive $(\epsilon >> 0)$ more oblate is the grain, and the more negative $(\epsilon << 0)$ more prolate is the grain.

\subsection*{Micromagnetic modelling}

\subsubsection*{Finite element meshes and magnetic properties}

In micromagnetic models, physical processes and their respective energies are evaluated mathematically through a continuous function, by assuming that at a given spatial volume the magnetisation is an average of discrete physical sources of magnetism, and that such volume is representative of the individual atomic behaviour of the averaged atoms \cite{Brown-1963,Hubert-1998,OConbhui-2018}. To solve the local energy minimum (LEM) for each of the iron oxides identified in our tomography, we rely on the open-sourced software package MERRILL \cite{OConbhui-2018}. This software uses linear tetrahedral finite elements to describe the geometry of a particle and solve the LEM stable states minimising the free energy of a magnetic system $E = \vec{H}_{\text{eff}} \cdot \vec{M}$, for which the total effective field $( \vec{H}_{\text{eff}} )$ is \cite{OConbhui-2018}: 

\begin{equation}\label{eq:Effective field}
\vec{H}_{\text{eff}} (\vec{M)}= \vec{H}_{\text{exg}} (\vec{M)} + \vec{H}_{\text{ans}} (\vec{M)} + \vec{H}_{\text{Zmn}} (\vec{M)} + \vec{H}_{\text{dmg}} (\vec{M)} 
\end{equation}

where $\vec{M}$ is the magnetisation vector, $\vec{H}_{\text{exg}}$ the exchange field, $\vec{H}_{\text{ans}}$ the anisotropy field, $\vec{H}_{\text{Zmn}}$ the Zeeman field, and $\vec{H}_{\text{dmg}}$ is the demagnetising field. 
We have used the software \emph{Cubit} to build the tetrahedral meshes of each particle (previously exported as a STL file). To ensure that our finite element meshes were fine enough to resolve the spatial variation of the magnetisation within the geometry of each grain we determined the maximum element size as dependent on their exchange length $\text{I}_\text{e}$ \cite{Rave-1998}. $\text{I}_\text{e}$ of magnetite $(\text{Fe}_{3}\text{O}_{4})$ is $\approx 9 \ \text{nm}$ at environmental conditions of temperature and pressure, and slightly larger than that near its \emph{Tc} \cite{OConbhui-2018}. The exchange length of maghemite $( \gamma \text{Fe}_{2}\text{O}_{3} )$ is slightly larger than magnetite's, so it is not an issue to assume $9 \ \text{nm}$ as a maximum element size for both mineralogies. Magnetic parameters of these minerals are set as in Table  \ref{tab: Magnetic_properties_table}.

\begin{table}[h]
  \centering
  \caption{Magnetic properties of the iron oxides used in our micromagnetic models at $20\ ^ \circ \text{C}$. $A_{ex}$ is the exchange constant, $K_1$ is the anisotropy constant, and $M_s$ is the magnetisation saturation.  }
  \label{tab: Magnetic_properties_table}
  \begin{tabular}{|c|c|c|c|c|c|}
    \hline
    \textbf{Mineral} & \textbf{Anisotropy} & \textbf{$A_{ex}$} & \textbf{$K_{1}$} & \textbf{$M_{s}$} & Ref \\
    \hline
    Magnetite $(Fe_{3}O_{4})$ & Cubic & $1.33 \cdot 10^{-11}$ & $-1.24 \cdot 10^{4}$ & $4.8 \cdot 10^5$ & Heider and Williams \cite{Heider-1988}  \\
    \hline
    Maghemite $(\gamma Fe_{2}O_{3})$ & Cubic & $1 \cdot 10^{-11}$ & $-4.6 \cdot 10^{3}$ & $3.8 \cdot 10^{5}$ & Dunlop and Özdemir \cite{Dunlop-1997} \\
    \hline

  \end{tabular}
\end{table}

\subsubsection*{Numerical calculations}

For any micromagnetic experiment performed through MERRILL, we can achieve a visualisation of the magnetic energy within the particles by performing simple mathematical operations to better illustrate the finite element solutions. We have achieved that through the open-sourced software \emph{Paraview} \cite{Ahrens-2005}. Firstly, we calculate the curl of the magnetisation $\vec{M} = M_{x}\hat{i}+M_{y}\hat{j}+M_{z}\hat{k} $, which is basically its vorticity $(\vec{\omega})$, as:

\begin{equation}\label{eq:vorticity}
\vec{\omega} = \nabla \times \vec{M}
\end{equation}

Sequentially, we calculate the projection of $\vec{\omega}$ onto $\vec {M}$, which is the normalised  helicity $(\vec {H})$ shown in our figures, as: 

\begin{equation}\label{eq:helicity}
\vec{H} = \frac{\vec{M} \cdot \vec{\omega}}{\lVert \vec{\omega} \rVert}
\end{equation}

When starting from random initial orientations, LEM calculations might reach slightly different products resulting from local minima. We calculate fifty (50) different LEM models for every grain at $20\ ^ \circ \text{C}$ and then harvest the states with the lowest minimised energies to reach the most stable magnetic domains (here called $r$). Upon categorising the states by their energies and magnetisation intensity/direction, we can determine the probability ($p_i$) associated with a particular state. This probability is expressed as a function of the state's energy and the system temperature, in the presence of other states, utilising a normalised Boltzmann probability distribution:

\begin{equation}\label{eq:Boltzmann_probability}
p_i = \frac{\exp\left(-(\varphi_i - \lambda_j)\right)}{\sum_{j}^{r}\exp\left(-(\varphi_j - \lambda_j)\right)}
\end{equation}

Here, $\varphi_i = \frac{E_{Mi}}{k_B \cdot T}$, where $E_{Mi}$ represents the magnetic energy (in J) of the state under consideration; $k_{B}$ is the Boltzmann constant $(1.3806 \times 10^{-23} \ \mathrm{m^2 \cdot kg \cdot s^{-2} \cdot K^{-1}})$; and the temperature of the system, denoted as $T$
(in K). Additionally, $\lambda_j$ is defined as $\min(\varphi_j)$. To ascertain the probability $(p_i)$ under varying temperatures, we conducted LEM minimisations for selected examples. In this process, each of the $r$-states served as the initial solutions. The temperatures ranged from $40-579 \ ^ \circ \text{C}$, with intervals of $20 \ ^ \circ \text{C}$ up to $500 \ ^ \circ \text{C}$, followed by intervals of $10 \ ^ \circ \text{C}$ up to $570 \ ^ \circ \text{C}$, and finally, $1 \ ^ \circ \text{C}$ increments up to $579 \ ^ \circ \text{C}$.

To assess the temporal stability of magnetic particles at environmental temperatures, we employed the Nudged–Elastic-Band (NEB) method \cite{Nagy-2017}. Initially, we sampled the minimum energy state of a given particle, and inverted the direction of magnetisation, resulting in two opposite (mirrored) domain states. The NEB method was then utilised to determine the magnetic energy (called energy barrier) required for the transition from one domain state to the other. For that, we simply gather the energy difference between the global maximum and minimum of the pathway. Subsequently, we have used this energy to calculate the relaxation time $(\tau)$ for each grain through the Néel–Arrhenius \cite{Neel-1955} equation as follows:

\begin{equation}\label{eq:Relaxation}
\tau = C \cdot \exp\left(\frac{Em}{k_B \cdot T}\right)
\end{equation}

where $C=10^{-9}  \ \text{s}$ is a frequency factor, and $T$ is fixed at $293.15 \ \text{K}$.

Magnetic hysteresis were evaluated from $-250 \ \text{mT}$ to $250 \ \text{mT}$  in $2 \ \text{mT}$ steps, while backfield curves were evaluated from $0$ to  $-250 \ \text{mT}$ using the same field steps, as this range covers the expected coercivity of such magnetic minerals. In our magnetic hysteresis simulations, we assess the magnetisation of a material under the influence of an applied magnetic field. The process initiates with saturation magnetisation (Ms) in a specific field direction, achieved at the maximum applied field. Subsequently, the field is gradually reduced to zero, resulting in the remanence of saturation (Mrs) at zero field. Following this, a magnetic field in the opposite direction (negative) is induced, where the field eventually causes the magnetisation to be zero (which is the coercivity, Bc), ultimately reaching the saturation point again. Backfield curves follow a similar process with a slight modification. The sample is first saturated, starting at zero field (Mrs). The field is then increased in steps, and before moving to a new step, it is decreased to zero. The remaining remanence is recorded, and as the remanence approaches zero, the coercivity of remanence (Bcr) is determined. To put it simply, magnetic hysteresis involves measuring magnetisation under the influence of a magnetic field, while backfield curves are influenced by a magnetic field, but the field is "turned off" during the actual "measurement" of the magnetisation. In our simulations of magnetic field-dependent experiments, the grain’s shape can create anisotropic properties due to easy/hard magnetisation axes. To account for this, we simulate different magnetic field orientations by sampling a field-dependent magnetic property \(F_i\) from random 3D coordinates (drawn from a Fibonacci sequence) on a unit sphere as:

\begin{equation}\label{eq:sphere averaging}
\iint_{s}{f(x,y,z)} \ ds\ \approx\ \sum_{i=1}^{20}\frac{F_i}{20}
\end{equation}

Here (Equation \ref{eq:sphere averaging}), \(f(x,y,z)\) evaluates the magnetic property \(F_i\), and the integral over the surface \(s\) is approximated by the sum over all samples. Because increasing the number of samples improves the approximation, we have run field-dependent experiments in twenty (\(\text{n}=20\)) random directions for every grain. 

\section*{Acknowledgements}

\subsection*{Funding}

This study was funded by grants $16/06114-6$, $21/00861-2$, $22/14100-6$ of the São Paulo Research Foundation (FAPESP). The opinions, hypotheses, and conclusions or recommendations expressed in this material are the responsibility of the authors and do not necessarily reflect the views of FAPESP. We extend our gratitude to Dra. Cilene de Medeiro and Dra. Ingrid Barcelos for their valuable assistance during the collection of SEM-EDS data (essential for selecting the area for FIB cutting) at the Microscopic Sample Laboratory, Brazilian Synchrotron Light Laboratory (LNLS, Proposal $20220316$). The authors gratefully acknowledge ScopeM for their support \& assistance in this work, especially to Anne Greet Bittermann, for the preparation of the FIB pillar. We acknowledge the Paul Scherrer Institut, Villigen, Switzerland for the provision of synchrotron radiation beamtime at beamline cSAXS of the SLS (Proposal $20221762$). U.D.B. expresses appreciation to the School of Geosciences at the University of Edinburgh for providing the computational resources essential for running the micromagnetic models presented in this paper. W.W. would like to acknowledge support from the Natural Environment Research Council (NERC) through grants $NE/V001388/1$, $NE/W006707/1$, and $NE/S011978/1$. D.G. acknowledges CNPq (Conselho Nacional de Desenvolvimento Científico e Tecnológico) through grant $310817/2020-0$. 

\subsection*{Author contributions}

U.D.B. performed sample preparation, data collection, processing, and analysis of tomography data, and conducted and interpreted the numerical models. W.W. contributed to running numerical models and supervising the project. R.I.F.T. conceived and supervised the overall project. A.D. was involved in collecting and processing tomography data. D.G. provided project supervision. All authors have actively contributed to the preparation and review of this manuscript.

\subsection*{Competing interests}

The authors declare no competing interests. 

\subsection*{Data,  Materials,  and  Software  Availability}

Besides the Supplementary File, all of the data related to this research is preserved at (https://doi.org/10.5281/zenodo.10837481) and can be publicly accessed from the same repository. The data is subdivided into folders, including the meshes used in our simulations (.pat files) and the MERRILL scripts, together with any other relevant file. We are also providing the averaged hysteresis and backfield curves of each grain. The results reported here were achieved through MERRILL \cite{OConbhui-2018}, an open-sourced code for micromagnetic modelling of tetrahedral finite element meshes. Installation, guide tutorials and courses are available at https://blogs.ed.ac.uk/rockmag/. 

\bibliography{sample}

\end{document}

\onehalfspacing

\title{Supplementary File: Coupling nanoscopic tomography and micromagnetic modelling to assess the stability of geomagnetic recorders}
\author[1,2,*]{Ualisson Donardelli Bellon}
\author[2,*]{Wyn Williams}
\author[1]{Ricardo Ivan Ferreira Trindade}
\author[3]{Ana Diaz}
\author[4]{Douglas Galante}
\affil[1]{University of São Paulo, Institute of Astronomy, Geophysics and Atmospheric Sciences (IAG), Department of Geophysics, São Paulo, 05360020, Brazil}
\affil[2]{University of Edinburgh, School of Geosciences, Edinburgh, EH9 3FE, Scotland}
\affil[3]{Paul Scherrer Institute, 5232 Villigen PSI, Switzerland}
\affil[4]{University of São Paulo, Institute of Geosciences, Department of Sedimentary and Environmental Geology, São Paulo, 05508080, Brazil}
\affil[*]{Correspondent authors: ualisson.bellon@usp.br, wyn.williams@ed.ac.uk }

\date{} % Remove the date

\maketitle

\section*{Contents}

This supplementary file provides additional insights into the methodology and results discussed in the main manuscript. Some of the detailed information on data acquisition is provided in the main manuscript and is not reiterated here. The figures are referenced in the main manuscript as \emph{Figure SX}, and the table presented here as \emph{Table S1}.

For access to codes and data, please visit \textbf{ZENODO}, where the files are publicly available. When using any content from this file, kindly refer to the main manuscript for proper citation.

\section{Ptychographic X-ray computed nano-tomography (PXCT) data acquisition and processing}

The X-ray illumination for PXCT was defined with a Fresnel zone plate (FZP) of 60~nm outer-most zone width and $200\,\mu$m diameter, in combination with a central stop of about $40\,\mu$m diameter, fabricated together with the FZP, and a sorting order aperture of $30~\mu$m. The FZP had locally displaced zones to produce an intentionally distorted illumination wavefront optimised for ptychography, see Ref.~\cite{odstrcil2019_uglyFZP}. The total flux of the beam was $5.0 \times 10^8$~photons/s. At the used energy of 6.2~keV, the FZP had a focal distance of 60.0~mm. The sample was placed at atmospheric pressure and room temperature 2.1~mm downstream of the focus, where the illumination had a size of about $7\,\mu$m. The sample was scanned following a Fermat spiral pattern~\cite{huang2014} with an average step size of $1\,\mu$m and a field of view of $37 \times 10\,\mu\text{m}^2$. At each scan position, coherent diffraction patterns were recorded with an Eiger~1.5M detector developed by the detector group at the Paul Scherrer Institute (PSI) in Villigen, Switzerland, with a pixel size of $75\,\mu$m. The detector was placed in a vacuum inside a flight tube to reduce air or any other window scattering and absorption. Ptychographic scans were repeated at 650 angular positions of the sample ranging from 0 to 180~deg in equally spaced angular steps. The tomographic acquisition was performed in a non-sequential manner, such that 8 successive series of full tomograms with a correspondingly larger angular step were acquired, in what is called a binary decomposition~\cite{kaestner2011}. Moreover, a few ptychographic scans were repeated at some angular positions, making a total of 656 projections recorded and used for tomography. With this acquisition approach, we could confirm that the sample did not change during acquisition due to radiation damage.

Ptychographic reconstructions were performed using the PtychoShelves package~\cite{wakonig2020}, with 300 iterations of a difference map algorithm~\cite{Thibault-2009} followed by 500 iterations of the maximum likelihood algorithm as a refinement~\cite{Thibault-2012}. In the reconstructions we used $400\times400$~pixels of the detector centred around the centre of the diffraction patterns, resulting in a reconstructed pixel size of 34.84~nm. The phase images from the reconstructions were aligned as described in the main text, and both the phase and absorption images were used for tomographic reconstruction by filter back projection, using a frequency cutoff of 1 and 0.25, respectively, resulting in 3D distributions of the electron density and $beta$, as described in the main text. The 3D resolution of the electron density dataset was estimated by Fourier shell correlation~\cite{vanHeel-2005}. In this approach, two subtomograms are computed, each with half of the tomographic projections, and their radially averaged correlation is compared to a threshold for a certain signal-to-noise ratio. Here we obtained a half-pitch resolution of 89~nm for the tomogram analysed in this manuscript, using the half-bit criterion~\cite{vanHeel-2005}. We note that this resolution represents the average over the entire specimen. It could be that the relatively small amount of high-density particles identified as magnetite could be imaged with a better resolution, given the high contrast between their density and that of their surrounding material.

The estimated dose imparted on the sample during the full acquisition of $5.6\times10^8$~Gy was estimated by measuring the number of photons absorbed by the sample with the Eiger detector and using an estimation of the sample mass based on the measured electron density distribution of the sample, as reported in~\cite{Bosch2023}.

\begin{figure}[h!]
\centering
\includegraphics[width=\linewidth]{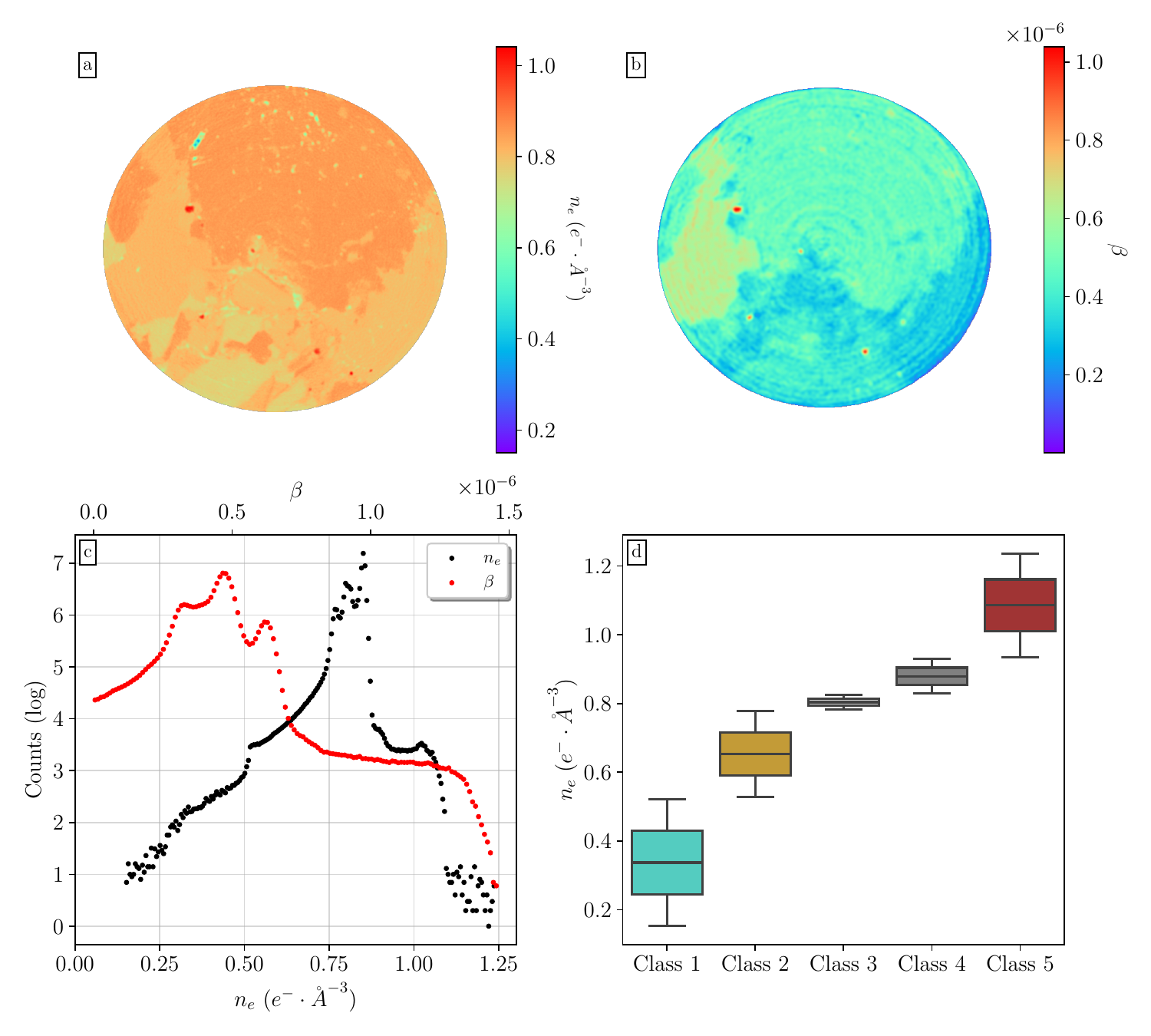}
\caption{Slices through the 3D reconstruction of the (a) electron density, $n_e$, and (b) $\beta$ tomograms by ptychographic X-ray computed tomography (PXCT). The colour map scheme was changed from greyscale to a rainbow to highlight the different materials better. Sections are about 18 $\mu \text{m}$ in diameter. The distribution of electron density and $\beta$ (the imaginary part of the full complex index of refraction) of all of the slices are shown in (c), removing the air contribution in the case of the $n_e$. d) After segmentation, five classes of distinct materials are identified based on their electron densities. A detailed description can be found in the \emph{Methods} section.}
\label{tomography_distribution}
\end{figure}

\begin{landscape}
\begin{table}[h]
  \centering
  \caption{Measured electron densities ($n_e$, $e^{-} / \text{Å}^{3}$) of the different segmented classes from the tomograms obtained through synchrotron-based ptychographic X-ray computed nano-tomography (nPXCT), along with the interpreted material and the corresponding calculated electron density. For the latter, we have approximated the ratio between molar mass and the number of electrons $(A / \phi)$ to $2$, a common assumption for most light atoms, with the notable exception of hydrogen \cite{Diaz-2015}. $Q_{75}$ is the upper quartile of the boxplot distribution of that given class, while $(\mu)$ is the median.  }
  \label{tab: Magnetic_properties_table}
  \begin{tabular}{|c|c|c|c|c|c|c|}
    \hline
    \textbf{ ED ($\mu$, measured)} & $Q_{75}$ | \textbf{Upper limit} &\textbf{Class} & \textbf{Material (interpreted)} & \textbf{Bulk density} & \textbf{ED (calculated)}  \\
    \hline
     $ 0.31 \pm 0.11$ &$0.43$|$0.52$& I & Bitumen & $1.02 - 1.03 \ g/cm^3$ \cite{Lapidus-2018} & $0.30 - 0.31$ \\
    \hline
    $0.65 \pm 0.08 $  &$0.71$|$0.78$& II & Smectite-Illite & $2.40 - 2.70 \ g/cm^3$ \cite{Totten-2002} & $0.69-0.72$\\
    \hline
    $0.80 \pm 0.01$  &$0.81$|$0.82$& III & Calcite & $2.71 \ g/cm^3$ \cite{mindat_calcite} & $0.81$  \\
    \hline
    $0.88 \pm 0.03$ &$0.90$|$0.92$& IV & Dolomite  & $2.76 - 2.84 \ g/cm^3$ \cite{mindat_dolomite} & $0.83 - 0.85$  \\
    \hline
   $ 1.08 \pm 0.09$  &$1.16$|$1.24$& V & Maghemite | Magnetite & $ 4.9$|$5.17 \ g/cm^3$ \cite{webmineral_maghemite, mindat_magnetite} & $1.46 - 1.55$ \\
    \hline

  \end{tabular}
\end{table}
\end{landscape}

\begin{figure}[ht]
\centering
    \includegraphics[width=9cm]{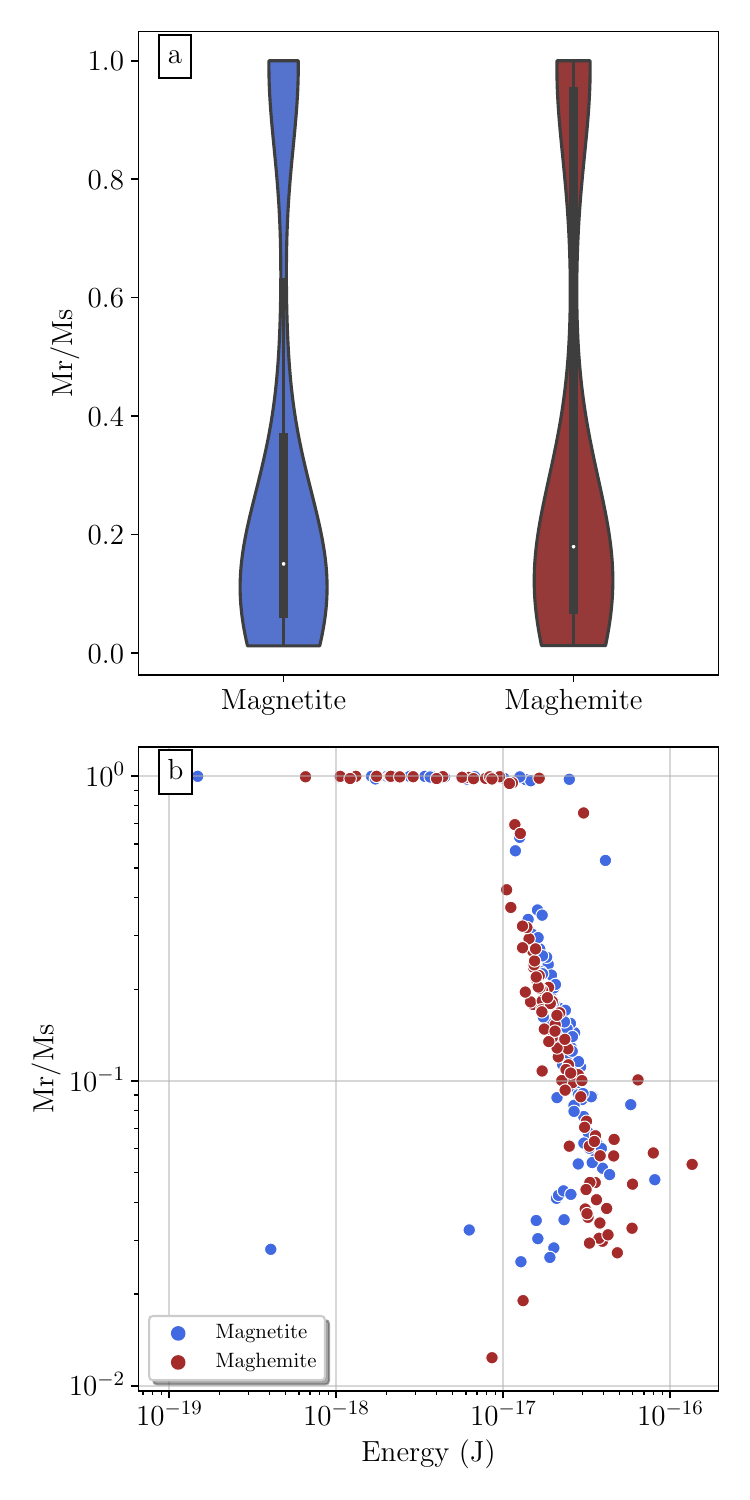}
\caption{Compilation of parameters acquired after running 50 Local Energy Minimum (LEM) models for each grain. These values refer to the minimum energy state found for each particle, for either magnetite (blue) or maghemite (brown). a) Violin plots showing the compilation of remanences (Mr/Ms) for each of the mineral species. The spreading from the centre of each plot regards the density distribution, whereas the boxplots within more clearly show the quartiles and their respective means (white dots). b) Final energies of each of these LEM states and the resulting remanence ratios, highlighting that both minerals occupy very similar domain states.}
\label{LEM_maghemite_magnetite}
\end{figure}

\begin{figure}[ht]
\centering
\includegraphics[width=\linewidth]{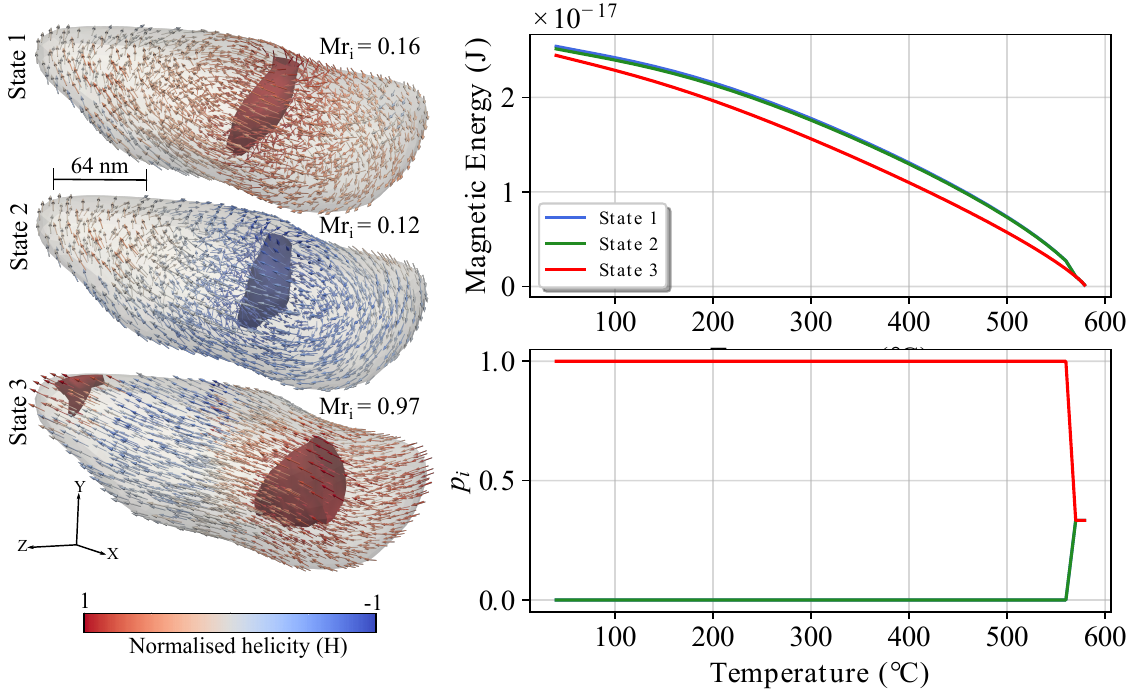}
\caption{The multiplicity of domain states and their statistical predominance across a range of temperatures. These LEM states exhibit very similar energies at environmental temperatures. Specifically, States 1 and 2 are both SV states, differing primarily in remanence magnitudes (as seen in $\text{Mr}_i$ values), while State 3 represents a flowering state with significantly higher remanence. As the temperature increases, States 1 and 2, with nearly identical magnetic energies, practically overlap up to $580^\circ \text{C}$, while State 3 becomes slightly less energetic (upper right plot). Despite this small difference in magnetic energy among the three states, it is sufficient, as determined by the Boltzmann probability density ($p_i$), to establish that State 3 will be the dominant state over the whole range of evaluated temperatures (lower right plot). We call this kind of vortex grain well-behaved (Group I), once it will probably succeed in both thermal demagnetisation and palaeointensity procedures. The interpretation of the magnetic structures follows the same as in the original manuscript.  }
\label{Grain_25_behaviour}
\end{figure}

\begin{figure}[ht]
\centering
\includegraphics[width=14.5cm]{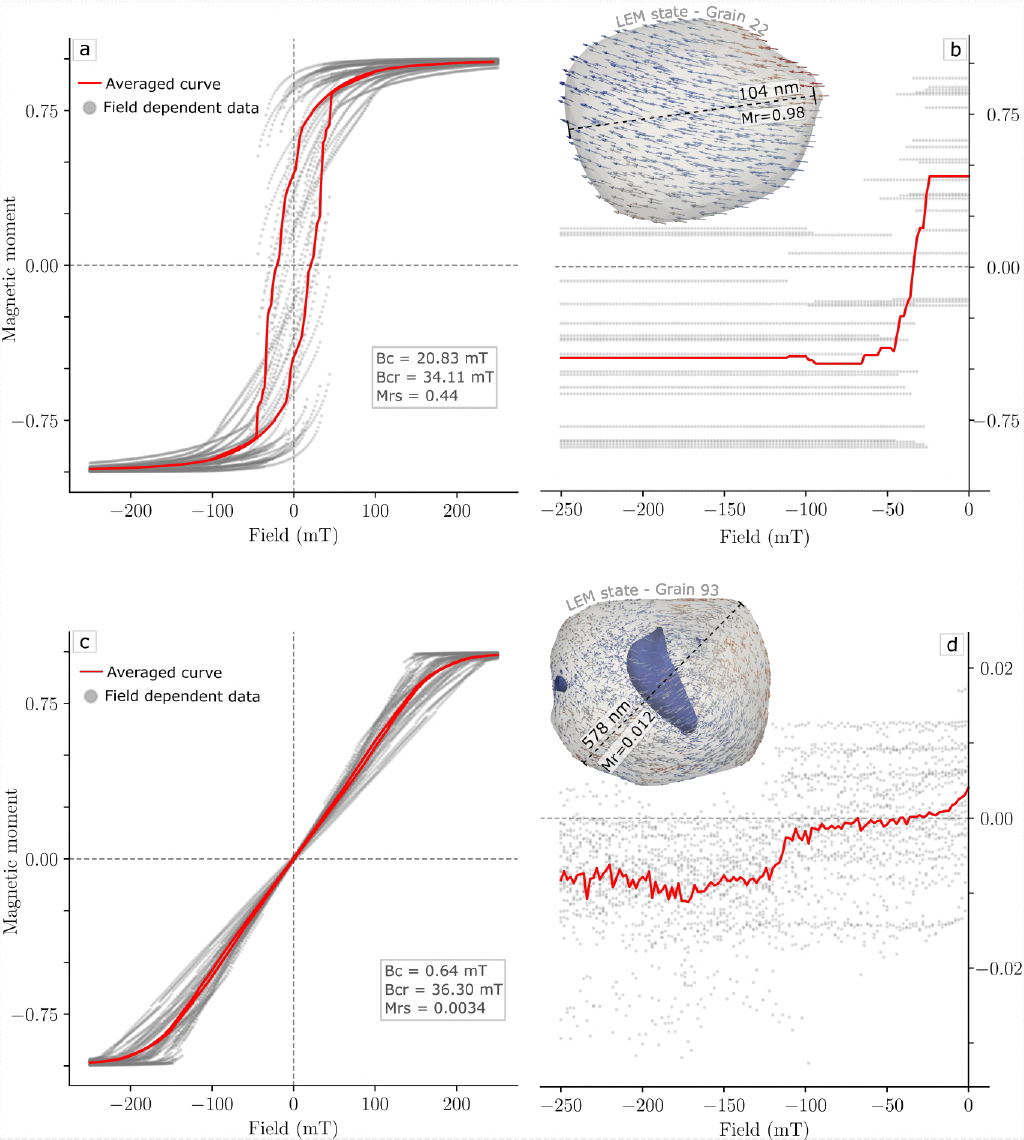}
\caption{The figures depict examples of magnetic hysteresis (a, c) and backfield curves (b, d). Each type of experiment involves data collection for 20 different field directions (grey points). The red lines represent the averaged curves obtained from these 20 different measurements. The grains showcased illustrate the remanence at zero field from a minimum energy LEM state, serving as a basis for comparison against the remanence of saturation from the magnetic hysteresis (Mrs). In panel (a), an archetype of magnetic hysteresis (Grain 22) is presented, characterised by a broad waist ($\text{Bc}=20.83 \ \text{mT}$), with Mrs being half the value of Mr (refer to the grain in b). The corresponding backfield curve (b) yields an averaged coercivity of remanence ($\text{Bcr}=34.11 \ \text{mT}$), which is comparable to the bulk coercivity (Bcr), as typically observed for grains with SD-like behaviour. Another example is illustrated in panel (c) for a subhedral particle (Grain 93) with a magnetic hysteresis featuring a thin waist ($\text{Bc}=0.64 \ \text{mT}$). However, as highlighted in the backfield curve (d), the coercivity of remanence is much higher ($\text{Bcr}=36.30 \ \text{mT}$), resulting in a high Bcr/Bc ratio.}
\label{Hysteresis_behaviour}
\end{figure}

\begin{figure}[ht]
\centering
\includegraphics[width=\linewidth]{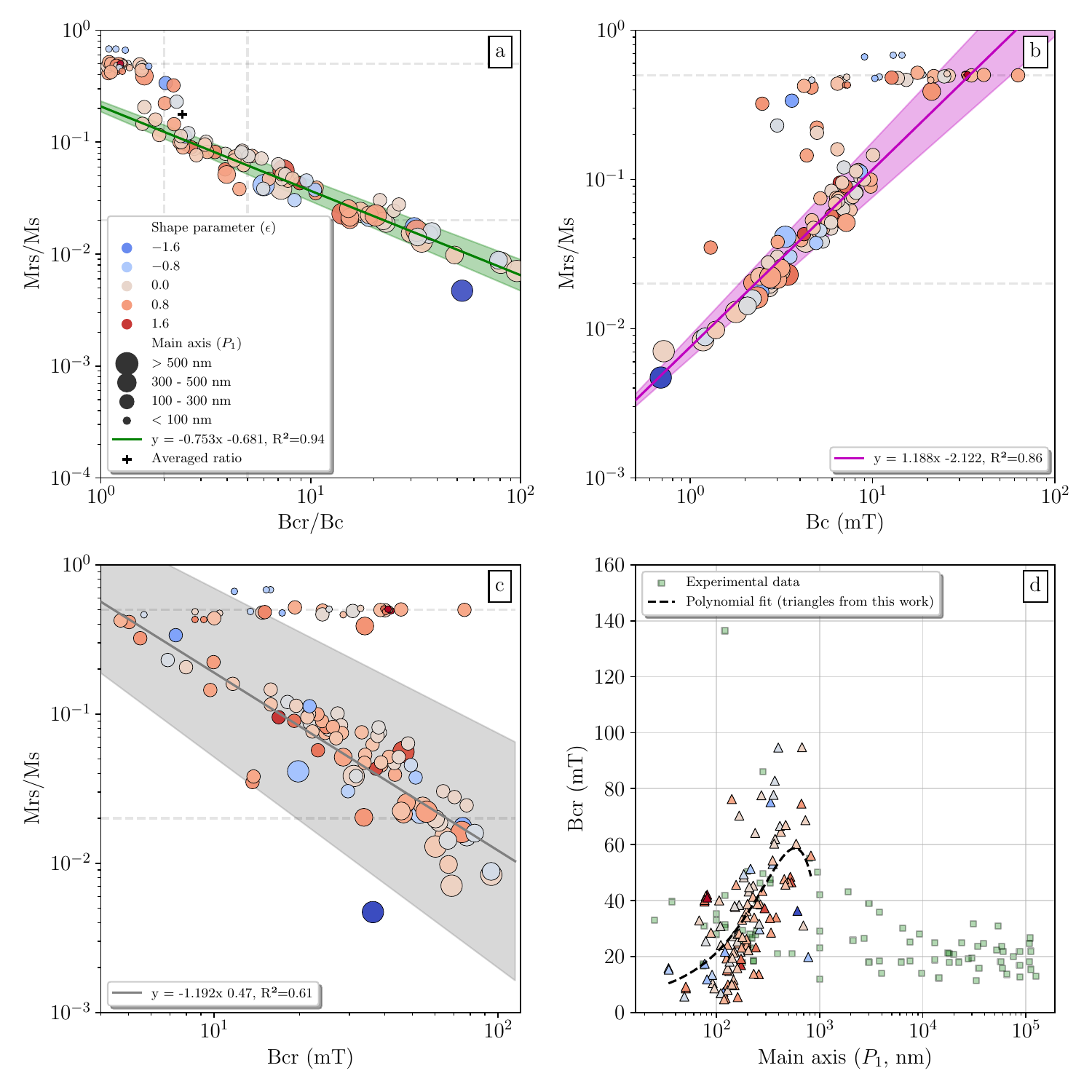}
\caption{The plots presented in this figure include: a) Bilogarithmic Day plot \cite{Day-1977, Dunlop-2002}; b) Neél's diagram \cite{Neel-1955}; and its c) sister plot; (d) Coercivity of remanence (Bcr) plotted against grain sizes, accompanied by experimental data (see references in the main manuscript). This plot and its interpretation mirror those presented in the main manuscript, with the distinction that these properties are calculated for maghemite.}
\label{Maghemite_parameters}
\end{figure}

\printbibliography